\documentclass{article}
\pdfoutput=1
\usepackage{graphicx}

\usepackage{longtable}

\usepackage[margin=1.0in]{geometry}
\usepackage{array}
\usepackage{verbatim}
\usepackage{color}
\usepackage{hyperref}
\hypersetup{
  colorlinks=blue,
  citecolor=blue,
  filecolor=blue,
  linkcolor=blue,
  urlcolor=blue
}

\usepackage{booktabs}
\usepackage{amsmath}
\usepackage{amssymb}
\usepackage{listings}

\usepackage{subfig}
\usepackage{wrapfig}

\newcommand{\Queso}{\texttt{QUESO}}
\newcommand{\MPI}{\texttt{MPI}}
\newcommand{\Gsl}{\texttt{GSL}}
\newcommand{\Libmesh}{\texttt{libMesh}}
\newcommand{\PyMC}{\texttt{PyMC}}
\newcommand{\emcee}{\texttt{emcee}}
\newcommand{\R}{\texttt{R}}
\newcommand{\Stan}{\texttt{Stan}}
\newcommand{\WinBUGS}{\texttt{WinBUGS}}
\newcommand{\DAKOTA}{\texttt{DAKOTA}}
\newcommand{\HDF}{\texttt{HDF5}}
\newcommand{\Boost}{\texttt{boost}}
\newcommand{\Teuchos}{\texttt{Teuchos}}
\newcommand{\Grvy}{\texttt{GRVY}}

\begin{document}
\title{The Parallel C++ Statistical Library for Bayesian Inference: QUESO}
\author{Damon McDougall, Nicholas Malaya, Robert D. Moser}
\maketitle

\abstract{
The Parallel C++ Statistical Library for the Quantification of
Uncertainty for Estimation, Simulation and Optimization, \Queso, is a
collection of statistical algorithms and programming constructs supporting
research into the quantification of uncertainty of models and their
predictions.  \Queso\ is primarily focused on solving statistical inverse
problems using Bayes's theorem, which expresses a distribution of possible
values for a set of uncertain parameters (the posterior distribution) in terms
of the existing knowledge of the system (the prior) and noisy observations of a
physical process, represented by a likelihood distribution.  The posterior
distribution is not often known analytically, and so requires computational
methods.  It is typical to compute probabilities and moments from the posterior
distribution, but this is often a high-dimensional object and standard
Reimann-type methods for quadrature become prohibitively expensive.  The
approach \Queso\ takes in this regard is to rely on Markov chain Monte Carlo (MCMC)
methods which are well suited to evaluating quantities such as probabilities
and moments of high-dimensional probability distributions.  \Queso's intended
use is as tool to assist and facilitate coupling uncertainty quantification to
a specific application called a \textit{forward problem}.  While many libraries
presently exist that solve Bayesian inference problems, \Queso\ is a
specialized piece of software primarily designed to solve such problems by
utilizing parallel environments demanded by large-scale forward problems.
\Queso\ is written in C++, uses \MPI, and utilizes libraries already available
to the scientific community.  Thus, the target audience of this library are
researchers who have solid background in Bayesian methods, are comfortable with
UNIX concepts and the command line, and have knowledge of a programming
language, preferably C/C++.
}


\section{Introduction} 

The Parallel C++ Statistical Library for the Quantification of Uncertainty for
Estimation, Simulation and Optimization, \Queso, is a collection of statistical
algorithms and programming constructs supporting research into the uncertainty
quantification (UQ) of models and their predictions. It has been designed with
three objectives: (a) to be sufficiently abstract in order to handle a large
spectrum of large-scale computationally intensive models; (b) to be extensible,
allowing easy creation of custom-defined objects; and (c) leverage parallel
computing through use of high-performance vector and matrix libraries.  Such
objectives demand a combination of an object-oriented design with robust
software engineering practices. \Queso\ is written in C++, uses \MPI, and
utilizes libraries already available to the scientific community.

The purpose of this book chapter is not to teach uncertainty
quantification methods, but rather to introduce the \Queso\ library so
it can be used as a tool to assist and facilitate coupling UQ to a
specific application (forward problem). Thus, the target audience of
this chapter is researchers who have solid background in Bayesian
methods, are comfortable with UNIX concepts and the command line, and
have knowledge of a programming language, preferably C/C++.

The rest of the document is organized as follows. Section \ref{sec:motivation}
has a brief discussion of statistical inverse problems, and in doing so,
provides the impetus behind the \Queso\ library.  Section \ref{sec:formulation}
then discusses the types of problems the library is designed to solve, as well
as introducing the notation used for the rest of this document. Several
illustrative examples, including the new infinite-dimensional capability, are
provided in section \ref{sec:examples} along with code snippets demonstrating
typical software call-patterns.  Section \ref{sec:extend} discusses how the
library design can easily be extended for bespoke user-defined random
variables, probability distribution functions, realizers.  Section
\ref{sec:design} discusses the design and internals of the library, as well as
providing a software snapshot of the current library status.  Finally, we
conclude by discussing several areas in which to focus future \Queso\
development efforts.

All of the examples in this document are present in the \Queso\ source tree of
the the latest release, 0.53.0.  Users should consult the website,
\url{libqueso.com}, for the latest news and source code.

This chapter builds on the 2012 paper that introduced the \Queso\ library
\cite{Prudencio2012} and the current \Queso\ user's manual
\cite{estacioquantification} by including a myriad of changes that have since
been incorporated into the library.

%
%

\section{Motivation}
\label{sec:motivation}

Statistical inverse problems using a Bayesian formulation model all quantities
as random variables, where probability distributions of the quantities capture
the uncertainty in their values.  The solution to the inverse problem is then
the probability distribution of the quantity of interest when all information
available has been incorporated in the model.  This (posterior) distribution
describes the degree of confidence about the quantity after the measurement has
been performed \cite{kaipio2005statistical}.

Thus, the solution to the statistical inverse problem is given by Bayes'
formula, which expresses the posterior distribution in terms of the prior
distribution and the data represented through the likelihood function.

For all but toy problems, the likelihood function has an open form and its
evaluation is highly computationally intensive.  Worse, simulation-based
posterior inference often requires a large number of these evaluations of the
forward model. Therefore, fast and efficient sampling techniques are desirable
for posterior inference.

It is often not straightforward to obtain explicit posterior point estimates of
the solution, since it usually involves the evaluation of a high-dimensional
integral with respect to a possibly non-smooth posterior distribution. In such
cases, an alternative integration technique is the Markov chain Monte Carlo
method where posterior moments may be estimated using the samples from a series
of (correlated) random draws from the posterior distribution.


\Queso\ is designed in an abstract way so that it can be used by any
computational model, as long as a likelihood function (in the case of
statistical inverse problems) and a quantity of interest (QoI) function (in the
case of statistical forward problems) is provided by the user application.

With this framework in mind, \Queso\ provides tools for both sampling
algorithms for statistical inverse problems, following Bayes' formula, and
statistical forward problems. It provides Markov chain Monte Carlo algorithms
using the Metropolis-Hasting acceptance ratio
\cite{Metropolis1953,Hastings1970}, these are this Multi-level Monte Carlo
\cite{Cheung2012} method and DRAM \cite{Haario2006}. \Queso\ is also capable 
of handling several chains in parallel computational environments.


\section{Alternatives to \Queso}

\Queso\ is certainly not the only quality statistical software library.  There
are many different libraries that can be used to solve Bayesian inference
problems.  \Queso\ is a specialized piece of software, primarily designed to
solve such problems utilizing the, often required, parallel environment
demanded by large-scale forward problems. This focus is simultaneously the
\Queso's greatest strength and weakness, depending on user's target problem.
For instance, \Queso\ would be less effective to use for serial problems than
several alternative libraries, as there is significant turnaround time from
learning how to build \Queso\ and link a custom forward code to it. In
instances where parallelization is not necessary and the forward problem is
relatively cheap to execute, there are good alternatives to \Queso.  We now
provide a simple survey of several other major libraries that we consider
useful for problems of Bayesian inference, along with a brief discussion some
unique strengths and weaknesses.

%
%

As discussed above, for inference problems that do not require parallelization,
serial libraries can be leveraged with less development.  An excellent example
of this is \PyMC\ \cite{Patil2010}. A modern software package, \PyMC\ is a
python-based software library for Bayesian estimation and MCMC.  Its strengths
lie in its flexibility and excellent post-processing, especially when coupled
with matplotlib \cite{Hunter2007}.  \emcee\ \cite{Foreman-Mackey2013} is
another python-based package, with a particular emphasis on Bayesian parameter
estimation. Both of these libraries are useful for rapid software prototyping
using serial inference problems.

%
%
On the other end of the spectrum, there are complete statistical software
languages. These are often more mature software projects which are capable of
much more general statistical computations than \Queso. However, these
languages are often weaker for specialized problems, because they are not as
well optimized for solving Bayesian inference problems, particularly at scale.
The ultimate example of this is \R\ \cite{Team2012}. \R\ is a free software
programming language and software environment for statistical computing and
graphics.  \R\ is arguably the most general and complete source of open-source
statistical packages in the world.  It is not limited to Bayesian techniques,
and has packages across a wide range of topics in statistics. However, it is
not easy to couple \R\ with other codebases (for the forward problem, for
instance).  Furthermore, while some packages supporting parallelism are now
being developed, the language is still primarily focused on serial
computations.  Another alternative is \Stan\ \cite{stan-software:2014}. \Stan\
is a probabilistic programming language written in C++ implementing full
Bayesian statistical inference.

Another major library is \WinBUGS\ \cite{Lunn2000}. \WinBUGS\ is statistical
software for Bayesian analysis using MCMC methods.
\WinBUGS\ is of particular historical importance, as it was one of the earliest
openly available MCMC libraries, with development starting the late 1980s.  It
is also unique in that it is developed for the Windows platform, instead of
Linux. It is also primarily based on the Gibbs sampler algorithm.

Finally, the \DAKOTA\ \cite{Adams2006} toolkit is a very general library
developed at Sandia National Laboratories, containing a vast array of
algorithms with applications to uncertainty quantification, optimization,
emulation, experimental design, prediction, and sensitivity analysis.  \DAKOTA\
is written in C++ and supported on Linux, OS X and on Windows and implements 20
years of advanced algorithms research.  Furthermore, given \DAKOTA's advanced
certainty propagation algorithms, the \Queso\ and \DAKOTA\ development teams are
working together to establish a seamless integration of \Queso's algorithms
into \DAKOTA\ to give users a matured and coupled forward and inverse UQ software
solution.


\section{Formulation}
\label{sec:formulation}

Here we give a rigorous description of the types of problems that \Queso\
solves.  This will crystallize both the terminology and notation in an attempt
to make everything in this chapter self-contained.

\subsection{The forward problem}

Here we set out the auspices under which we will operate.  We make two
high-level assumptions: 1) we have access to a set of observations of some
physical phenomenon; and 2) we have a mathematical model that attempts to model
the observed physical phenomenon.  Ensuring that the mathematical
model is \emph{valid} is an exercise left to the reader.  We will denote the
observed data by $y$, and the mathematical model by $\mathcal{G}$.  The model
will certainly depend on various parameters, and we call the process of mapping
these parameters to model output the \emph{forward problem}.  In many physical
engineering applications, the forward problem is expensive and may involve the
solution of a set of partial differential equations.

\subsection{The inverse problem}

In the subsection above, we described the forward problem.  It may be the case
that the mathematical model in the forward problem may depend on some parameters
that are unknown and that we wish to estimate.  We will refer to these unknown
parameters as $\theta$. The process of estimating $\theta$ given observations
goes by many names, but is generally referred to as the \emph{inverse problem}.
There are several frameworks for solving inverse problems.  We will focus only
on the \emph{Bayesian framework}, which we rigorously describe now.

As noted above, we are given a set of observations $y$.  This dataset is
corrupted by errors made during the experiment.  These errors could be human
errors, equipment errors, or errors in the setup of the experimental scenario.
In complete generality, it is difficult to say with certainty what statistical
distribution these errors follow.  In a lot of experimental cases, however,
a Gaussian distribution with some, perhaps unknown, variance is quite a
reasonable characterization.

The unknown parameters themselves might have some inherent constraining
property.  For example, if the unknown parameter were a concentration of a
contaminant underground then it is not possible for this unknown parameter to
be negative.  The constraint varies depending on the physical domain, but it is
rarely the case one knows \emph{nothing} about the unknown parameters.  This
information can be translated to constraints on a prior distribution.

To regroup, we have a statistical distribution governing the behavior of the
experimental errors given the unknown parameters, $\mathbb{P}(y | \theta)$.  We
also have some prior distribution on the unknown parameters
$\mathbb{P}(\theta)$.  The Bayesian solution to the inverse problem of finding
$\theta$ is the distribution of $\theta$ given $y$, $\mathbb{P}(\theta | y)$.
By Bayes's rule, this can be written as follows,
\begin{equation*}
  \mathbb{P}(\theta | y) \propto \mathbb{P}(y | \theta) \mathbb{P}(\theta).
\end{equation*}
The left-hand side is referred to as the posterior distribution.  The
right-hand side is the product of the likelihood distribution and the prior
distribution.  \Queso\ solves the Bayesian inverse problem by providing samples
that are distributed according to the posterior distribution using 
Markov chain Monte Carlo.  This chapter does not provide the details of how
MCMC works, but refer the reader to the expansive body of available literature
on the topic cited throughout this work.

\subsection{Prediction}

The prediction step in the Bayesian framework is that of estimating some
quantity $\mathcal{Q}(\theta)$ dependent on the unknown parameters.  This is
usually referred to as a \emph{statistical forward problem}.  \Queso\ is
equipped to solve statistical forward problems, but throughout this chapter we
will focus mainly on the statistical inverse problem.

\section{Examples}
\label{sec:examples}

\subsection{A template example}

Here we walk through a template example.  This template should be general
enough to serve as a good starting point for most Bayesian inverse problems.
Before we step through the example, here it is in its entirety:
\begin{verbatim}
#include <queso/GslVector.h>
#include <queso/GslMatrix.h>
#include <queso/UniformVectorRV.h>
#include <queso/StatisticalInverseProblem.h>
#include <queso/ScalarFunction.h>
#include <queso/VectorSet.h>

template<class V = QUESO::GslVector, class M = QUESO::GslMatrix>
class Likelihood : public QUESO::BaseScalarFunction<V, M>
{
public:

  Likelihood(const char * prefix, const QUESO::VectorSet<V, M> & domain)
    : QUESO::BaseScalarFunction<V, M>(prefix, domain)
  {
    // Setup here
  }

  virtual ~Likelihood()
  {
    // Deconstruct here
  }

  virtual double lnValue(const V & domainVector, const V * domainDirection,
      V * gradVector, M * hessianMatrix, V * hessianEffect) const
  {
    // 1) Run the forward code at the point domainVector
    //    domainVector[0] is the first element of the parameter vector
    //    damainVector[1] is the second element of the parameter vector
    //    and so on
    //
    // 2) Compare to data, y
    //    Usually we compute something like
    //    || MyModel(domainVector) - y ||^2 / (sigma * sigma)
    //
    // 3) Return below

    double misfit = 0.0;

    return -0.5 * misfit;
  }

  virtual double actualValue(const V & domainVector, const V * domainDirection,
      V * gradVector, M * hessianMatrix, V * hessianEffect) const
  {
    return std::exp(this->lnValue(domainVector, domainDirection, gradVector,
          hessianMatrix, hessianEffect));
  }

private:
  // Maybe store the observed data, y, here.
};

int main(int argc, char ** argv) {
  MPI_Init(&argc, &argv);

  // Step 0 of 5: Set up environment
  QUESO::FullEnvironment env(MPI_COMM_WORLD, argv[1], "", NULL);

  // Step 1 of 5: Instantiate the parameter space
  QUESO::VectorSpace<> paramSpace(env,
      "param_", 1, NULL);

  double min_val = 0.0;
  double max_val = 1.0;

  // Step 2 of 5: Set up the prior
  QUESO::GslVector paramMins(paramSpace.zeroVector());
  paramMins.cwSet(min_val);
  QUESO::GslVector paramMaxs(paramSpace.zeroVector());
  paramMaxs.cwSet(max_val);

  QUESO::BoxSubset<> paramDomain("param_", paramSpace, paramMins, paramMaxs);

  // Uniform prior here.  Could be a different prior.
  QUESO::UniformVectorRV<> priorRv("prior_", paramDomain);

  // Step 3 of 5: Set up the likelihood using the class above
  Likelihood<> lhood("llhd_", paramDomain);

  // Step 4 of 5: Instantiate the inverse problem
  QUESO::GenericVectorRV<> postRv("post_", paramSpace);

  QUESO::StatisticalInverseProblem<> ip("", NULL, priorRv, lhood, postRv);

  // Step 5 of 5: Solve the inverse problem
  QUESO::GslVector paramInitials(paramSpace.zeroVector());

  // Initial condition of the chain
  paramInitials[0] = 0.0;
  paramInitials[1] = 0.0;

  QUESO::GslMatrix proposalCovMatrix(paramSpace.zeroVector());

  for (unsigned int i = 0; i < 2; i++) {
    // Might need to tweak this
    proposalCovMatrix(i, i) = 0.1;
  }

  ip.solveWithBayesMetropolisHastings(NULL, paramInitials, &proposalCovMatrix);

  MPI_Finalize();

  return 0;
}
\end{verbatim}
Notice that this template example is fairly short, weighing in at roughly 100
lines of boilerplate C++ code.  Incorporating a specific physical model into
the likelihood will certainly increase the size of the statistical application.
In the meantime, we will walk through the boilerplate setup that will be
common to many use-cases.

We will start with the \texttt{main} function.  This is where most of the setup
takes place.  Firstly, since \Queso\ uses \MPI, we must call the
\texttt{MPI\_Init} function before using any of the classes in \Queso.  The
next line,
\begin{verbatim}
QUESO::FullEnvironment env(MPI_COMM_WORLD, argv[1], "",
    NULL);
\end{verbatim}
sets up the \Queso\ environment.  The constructor parameters are, in order, an
\MPI\ communicator and could be a custom sub-communicator; the filename of a
\Queso\ input file; a prefix, if a different from the default is desired, for
input file options specific to the \Queso\ environment; and an optional
\texttt{EnvOptionsValues} object so that the user can set environment options
programmatically. The next thing we do is define the dimension of the state
space by created a object representing a vector space:
\begin{verbatim}
QUESO::VectorSpace<> paramSpace(env, "param_", 1, NULL);
\end{verbatim}
In this particular example, the dimension of the state space is 1.  The
constructor parameters here are the \Queso\ environment; a prefix, if a
different from the default is desired, for input file options specific to this
parameter space object; and a vector of strings to name components of the
vectors belonging to this vector space.  Now we are in a position to set up the
domain of the statistical inverse problem.  \Queso\ only supports box domains
but the bounds for the box may be arbitrary.  We store the bounds for the
domain in \texttt{GslVector} objects like so:
\begin{verbatim}
QUESO::GslVector paramMins(paramSpace.zeroVector());
paramMins.cwSet(min_val);
QUESO::GslVector paramMaxs(paramSpace.zeroVector());
paramMaxs.cwSet(max_val);
\end{verbatim}
Here \texttt{min\_val} and \texttt{max\_val} will be specific to the user's
problem.  The box domain uses these bounds and is constructed as follows:
\begin{verbatim}
QUESO::BoxSubset<> paramDomain("param_", paramSpace,
    paramMins,
    paramMaxs);
\end{verbatim}
We have finished setting up the domain of the statistical inverse problem.
Recall the ingredients we need for a well-posed statistical inverse problem; a
prior distribution and a likelihood distribution.  \Queso\ supports many
statistical distributions that can all be used as a prior, and the user may
choose to implement their own prior distribution if (see \ref{sec:extend}) such
customization is needed.  The following line creates an object representing a
uniform random variable:
\begin{verbatim}
QUESO::UniformVectorRV<> priorRv("prior_",
    paramDomain);
\end{verbatim}
This object contains all the necessary information to fully define a uniformly
distributed random variable.  Namely, its probability density function, and
mechanisms by which one can make draws with this density.  The second
ingredient needed for a statistical inverse problem is the definition of a
likelihood distribution, and this is done now:
\begin{verbatim}
Likelihood<> lhood("llhd_", paramDomain);
\end{verbatim}
This line may look different to the one for your specific application, as it is
intended to interact with a specific physical model.  The \texttt{Likelihood}
class is a custom-defined class.  We will come back to the full
\texttt{Likelihood} class in sections \ref{sec:balldrop} and
\ref{sec:define-lhd} explain how it is implemented.  For now, we will continue
with the setup of the inverse problem, and all the necessary code needed to
initialize the sampling.  We construct a placeholder object that represents a
posterior random variable:
\begin{verbatim}
QUESO::GenericVectorRV<> postRv("post_", paramSpace);
\end{verbatim}
\Queso\ will operate on this object during the sampling.  After \Queso\ has
finished its sampling, this object is then available to you for
post-processing.  Next, we pass the prior, likelihood and posterior over to
the \texttt{StatisticalInverseProblem} class like so:
\begin{verbatim}
QUESO::StatisticalInverseProblem<> ip("", NULL,
    priorRv, lhood, postRv);
\end{verbatim}
We are now ready to finalize the setup of the inverse problem.  We do this by
giving \Queso\ an initial condition for the sampler:
\begin{verbatim}
QUESO::GslVector paramInitials(
    paramSpace.zeroVector());
paramInitials[0] = 0.0;
paramInitials[1] = 0.0;
\end{verbatim}
We also give \Queso\ an initial covariance matrix:
\begin{verbatim}
QUESO::GslMatrix proposalCovMatrix(
    paramSpace.zeroVector());
for (unsigned int i = 0; i < 1; i++) {
  proposalCovMatrix(i, i) = 0.1;
}
\end{verbatim}
The closer this matrix is to the covariance between parameters under the
posterior measure, the better the Markov chain will perform.  Providing a bad
initial covariance does not change the posterior distribution in the limit of
infinite samples.  Finally, we begin sampling with the following call:
\begin{verbatim}
ip.solveWithBayesMetropolisHastings(NULL,
    paramInitials, &proposalCovMatrix);
\end{verbatim}

\subsection{Defining the likelihood distribution}
\label{sec:define-lhd}

As can be observed in the example illustrated above, the user must pass a
likelihood to \Queso.  \Queso\ expects, as a likelihood, anything that
subclasses the \texttt{BaseScalarFunction} abstract base class.  This base
class has two pure virtual functions that must be implemented in any subclass.
These functions are \texttt{lnValue()} and \texttt{actualValue()}.  The
function \texttt{lnValue} take a number of parameters, the most important of
which is \texttt{const V \& domainVector}.  When the user implements this
function it should return the natural logarithm of the likelihood distribution
evaluated the point \texttt{domainVector}.  A concrete example of this can be
seen in the next subsection.  The function \texttt{actualValue} should return
exactly the likelihood distribution evaluted at the point
\texttt{domainVector}.  For most practical applications, this function will
usually just return \texttt{std::exp} of \texttt{lnValue}, but the user has
the freedom to implement a more optimized computation if one is needed.

A typical Gaussian likelihood distribution will look something like this
\begin{verbatim}
template<class V, class M>
double
Likelihood<V = QUESO::GslVector,
    M = QUESO::GslMatrix>::lnValue(
    const V & domainVector,
    const V * domainDirection, V * gradVector,
    M * hessianMatrix, V * hessianEffect) const
{
  double misfit = 0.0;
  unsigned int vec_len = domainVector.sizeLocal()

  for (unsigned int i = 0; i < vec_len; i++) {
    misfit += domainVector[i] - observation[i];
  }

  return -0.5 * misfit;
}
\end{verbatim}

To avoid numerical problems computing the acceptance probability in an MCMC
algorithm, \Queso\ will call \texttt{lnValue} instead of \texttt{actualValue}
to do the accept-reject step in log space.

\subsection{Ball drop example}
\label{sec:balldrop}

\newcommand{\bv}[1]{\ensuremath{\mbox{\boldmath$ #1 $}}}
\newcommand{\chainsizeresults}{20000}

This section presents an example of how to use \Queso\ as an application that
solves a statistical inverse problem (SIP) and a statistical forward problem
(SFP), where the solution of the former serves as input to the later. This
example will use the canonical ``Ball Drop'' problem, a standard problem in
uncertainty quantification.  The objective of the SIP is to infer the the
acceleration due to gravity on an object in free fall near the surface of the
Earth. During the SFP, the distance traveled by a projectile launched at a
given angle and altitude is calculated using the calibrated magnitude of the
acceleration of gravity gathered during the SIP. As expressed in the section
\ref{sec:formulation}, both the inference and forward problem will be performed
using a Bayesian methodology, and so, the resulting quantities of interest
(QoIs) will be expressed as probability distributions.

\subsection{Statistical Inverse Problem}
\label{sec:gravity-ip}



A deterministic mathematical model for the vertical motion of an object in free
fall near the surface of the Earth is given by,
\begin{equation}
  \label{eq:gravity01}
  h(t) = -\frac{1}{2} g t^2 + v_0 t + h_0.
\end{equation}
where, $v_0$ [$m/s$] is the initial velocity, $h_0$ [$m$] is the initial
altitude, $h(t)$ [$m$] is the altitude with respect to time, $t$ [$s$] is the
elapsed time, and $g$ [$m/s^2$] is the magnitude of the acceleration due to
gravity (the parameter which cannot be directly measured and will be
statistically inferred).

This model is an expression of a high fidelity model, Newton's second law of
motion.  However, the model is imperfect, as it does not account resistive
force of air resistance, for example.

\subsubsection{Experimental data}

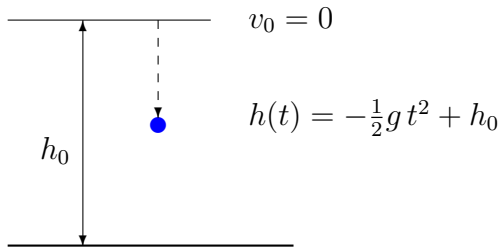
\begin{figure}[!ht]
\centering
\setlength{\unitlength}{4144sp}%
\begingroup\makeatletter\ifx\SetFigFont\undefined%
\gdef\SetFigFont#1#2#3#4#5{%
  \reset@font\fontsize{#1}{#2pt}%
  \fontfamily{#3}\fontseries{#4}\fontshape{#5}%
  \selectfont}%
\fi\endgroup%
\begin{picture}(1754,1477)(249,-803)
{\color[rgb]{0,0,1}\thinlines
\put(1171,-61){\circle*{90}}
}%
{\color[rgb]{0,0,0}\multiput(1171,569)(0.00000,-106.36364){6}{\line( 0,-1){ 53.182}}
\put(1171,-16){\vector( 0,-1){0}}
}%
\thicklines
{\color[rgb]{0,0,0}\put(271,-781){\line( 1, 0){1710}}
}%
\thinlines
{\color[rgb]{0,0,0}\put(271,569){\line( 1, 0){1215}}
}%
{\color[rgb]{0,0,0}\put(721,569){\vector( 0, 1){  0}}
\put(721,569){\vector( 0,-1){1350}}
}%
\put(465,-264){\makebox(0,0)[lb]{\smash{{\SetFigFont{12}{14.4}{\rmdefault}{\mddefault}{\updefault}{\color[rgb]{0,0,0}$h_0$}%
}}}}
\put(1711,524){\makebox(0,0)[lb]{\smash{{\SetFigFont{12}{14.4}{\rmdefault}{\mddefault}{\updefault}{\color[rgb]{0,0,0}$v_0=0$}%
}}}}
\put(1711,-61){\makebox(0,0)[lb]{\smash{{\SetFigFont{12}{14.4}{\rmdefault}{\mddefault}{\updefault}{\color[rgb]{0,0,0}$h(t)=-\frac{1}{2} g\,t^2+h_0$}%
}}}}
\end{picture}%
\vspace*{-8pt}
\caption{An object falls from altitude $h_0$ with zero initial velocity ($v_0=0$).}
\label{fig:free_fall}
\end{figure}

The experimental data will be generated from an identical object falling from
several different heights, each with zero initial velocity (See Figure
\ref{fig:free_fall}). We collect data, $\mathbf{y}$, of the time taken for the
ball to impact the ground starting from various different initial heights.
Each experimental observation error is treated as Gaussian with some known mean
and variance standard deviation, $\sigma$. The error is a result of measurement
uncertainties, such as estimates of the actual height the object was dropped
from, the human error introduced by operating a stop watch for time
measurement, and any other possible sources of error.  The actual observation
values can be found in the accompanying source code that will follow shortly.

\subsubsection{The prior, likelihood and posterior}

In Bayesian inference, the prior probability signifies the modeler's
expectation of the result of an experiment before any data is provided. In this
problem, a prior must be provided for the parameter $g$.  Near the surface of
the Earth, an object in free fall in a vacuum will accelerate at approximately
$9.8 m/s^2$, independent of its mass. For this gravitational inference problem,
we will place a uniform prior on our unknown variable $\theta$, over the
interval [8,11]:
\begin{equation}
  \label{eq-g-prior}
  \mathbb{P}(\theta) = \mathcal{U}(8,11).
\end{equation}

We select a Gaussian likelihood function that assigns greater probabilities to
parameter values that result in model predictions close to the data:
\begin{equation}
  \label{eq:like02}
  \mathbb{P}(\mathbf{y} | \theta) \varpropto
  \exp \left( -\frac12
  \left( \mathcal{G}(\theta) - \mathbf{y}\right)^T
  \mathbf{C}^{-1}
  \left( \mathcal{G}(\theta) - \mathbf{y} \right)
  \right),
\end{equation}
where $\mathbf{C}$ is a given covariance matrix, $\mathbf{y}$ is the
experimental data, and $\mathcal{G}(\theta)$ is the model output.

Using the deterministic model for the acceleration of gravity (Eqn.
\ref{eq:gravity01}) with no initial velocity, the observations $\mathbf{y}$,
and equation \eqref{eq:like02}, we have:
\begin{equation}
  \label{eq:like03}
  \theta \stackrel{\text{\small{def.}}}{=} g,
  \quad
  \mathcal{G}(\theta)=
  \left[
    \begin{array}{c}
      \sqrt{\dfrac{2 h_1}{g}}\\
      \sqrt{\dfrac{2 h_2}{g}}\\
      \vdots\\
      \sqrt{\dfrac{2 h_{n_d}}{g}}
    \end{array}
  \right],
  \quad
  \mathbf{y} =
  \left[
    \begin{array}{c}
      t_1    \\
      t_2    \\
      \vdots \\
      t_{n_d}
    \end{array}
  \right],
  \quad
  \mathbf{C} =
  \left[
    \begin{array}{cccc}
      \sigma^2_1 & 0	        & \cdots & 0 \\
      0          & \sigma^2_2 & \cdots & 0 \\
      \vdots     & \vdots     & \ddots & 0 \\
      0          & 0          & \cdots & \sigma^2_{n_d}
    \end{array}
  \right],
\end{equation}
where $n_d=14$ is the number of observations. We now invoke Bayes's formula in
order to obtain the posterior PDF $\mathbb{P}(\theta | \mathbf{y})$,
\begin{equation}
  \label{eq-Bayes-g}
  \mathbb{P}(\theta | \mathbf{y}) \varpropto \mathbb{P}(\mathbf{y} | \theta)
  \mathbb{P}(\theta).
\end{equation}

\subsection{Statistical forward problem}

\begin{figure}[!ht]
\centering
\includegraphics[scale=1]{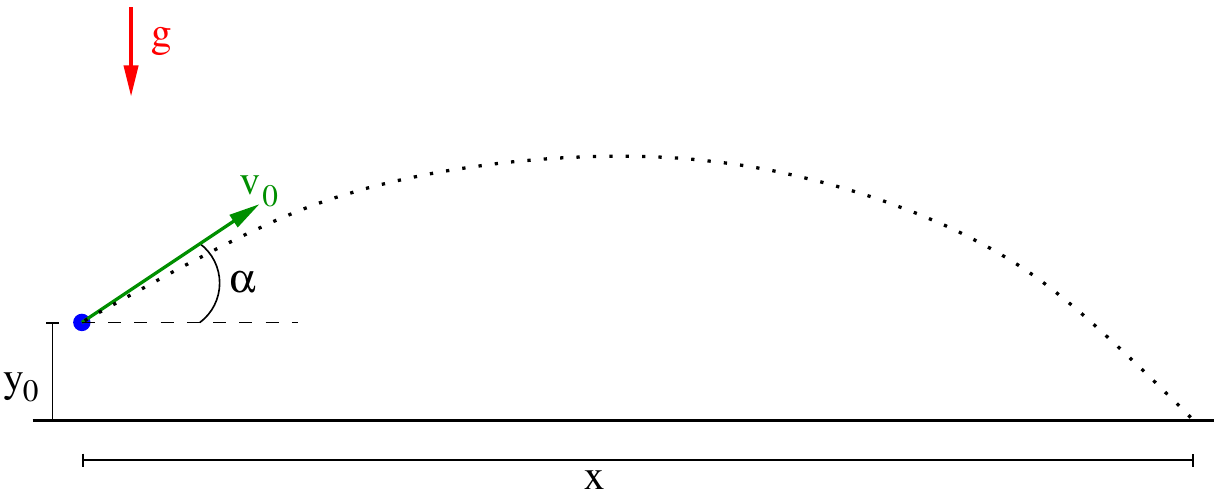}
\vspace*{-8pt}
\caption{Object traveling with projectile motion. }
\label{fig:projectile}
\end{figure}

Projectile motion refers to the motion of an object projected into the air at
an angle, e.g. a soccer ball being kicked, a baseball being thrown, or an
athlete long-jumping. In the absence of a propulsion system and neglecting air
resistance, the only force acting on the object is proportional to a constant
gravitational acceleration $g$. A deterministic two-dimensional mathematical
model for the vertical motion of an object projected from near the surface of
the Earth is given by,
\begin{align}
  \label{eq:fwd01}
  v_x &= v_{0x}, \\
  v_y &= v_{0y} - gt, \\
    x &= v_{0x}t, \\
    h &= h_0 + v_{0y}t - \frac{1}{2} g t^2,
\end{align}
where $h_0$ is the initial height, $x=x(t)$ is the distance traveled by the
object, $\bv{v_0}=(v_{0x},v_{0y})$ is the initial velocity, $v_{0x} = v_{0}
\cos(\alpha)$, $v_{0y} = v_{0} \sin(\alpha)$, and $v_0=\|\bv{v_0}\|^2$.
Figure \ref{fig:projectile} displays the projectile motion of an object in
these conditions.

In this example, we assume that $h_0,\alpha\text{ and }v_0$ are all known, with
$h_0 = 0$, $\alpha = \pi/4$, $v_0 = 5$ and $g$ is the result of the SIP
described in section \ref{sec:gravity-ip}.

Since the result of the SIP is a PDF on $g$, the output of the mathematical
model (\ref{eq:fwd01}) will be a random variable, and our forward problem
result will also be statistical in nature.

\subsubsection{The input random variable, QoI, and output random variable}

The input for the statistical forward problem is the random variable $g$, the
acceleration of gravity.  This is the solution (posterior PDF) of the inverse
problem described in Section \ref{sec:gravity-ip}.  The QoI for this example is
the distance $x$ traveled by an object in projectile motion.

Combining the expressions in Equation \eqref{eq:fwd01} and rearranging them,
the QoI function for $x$ is:
\begin{equation}
  \label{eq:fp_deterministic}
  x = \dfrac{ v_0 \cos \alpha }{g} \left( v_0 \sin \alpha + \sqrt{(v_0 \sin \alpha)^2 + 2g \, y_0} \right).
\end{equation}
Here $x$ is the distance traveled and our quantity of interest (QoI).

\subsection{Example code}
\label{sec:gravity_code}

The source code for the SIP and the SFP is composed of several files.  Three of
them are common for both problems: \texttt{gravity\_main.C, gravity\_compute.h}
and \texttt{gravity\_compute.C}; they combine both problems and use the
solution of the SIP (the posterior PDF for the gravity) as an input for the
SFP.  We present only the statistical inverse problem here.  The forward
problem is very similar to the inverse problem and the user is encouraged to
visit the source tree (\url{https://libqueso.com}) for the full treatment.

The files common to the inverse (and forward) problem are in Listings
\ref{code:gravity_main} and \ref{code:gravity_compute_C}.  Two files
specifically handle the SIP: \texttt{gravity\_likelihood.h} and
\texttt{gravity\_likelihood.C}. These are displayed in Listings
\ref{code:gravity_like_h} and \ref{code:gravity_like_C}.

\begin{lstlisting}[caption=File \texttt{gravity\_main.C.},
                 label=code:gravity_main]
#include <gravity_compute.h>

int main(int argc, char* argv[])
{
  // Initialize QUESO environment
  MPI_Init(&argc,&argv);
  QUESO::FullEnvironment* env =
    new QUESO::FullEnvironment(MPI_COMM_WORLD,argv[1],"",NULL);

  // Call application
  computeGravityAndTraveledDistance(*env);

  // Finalize QUESO environment
  delete env;
  MPI_Finalize();

  return 0;
}
\end{lstlisting}

\begin{lstlisting}[caption={File \texttt{gravity\_compute.C}. The first part of
                 the code (lines 4--44) handles the statistical forward
                 problem, whereas the second part of the code (lines 53--76)
                 handles the statistical forward problem.\\},
                 label=code:gravity_compute_C,numbers=left]
void computeGravityAndTraveledDistance(const QUESO::FullEnvironment& env) {
  // Statistical inverse problem (SIP): find posterior PDF for 'g'

  // SIP Step 1 of 6: Instantiate the parameter space
  QUESO::VectorSpace<QUESO::GslVector,QUESO::GslMatrix> paramSpace(env,
      "param_", 1, NULL);

  // SIP Step 2 of 6: Instantiate the parameter domain
  QUESO::GslVector paramMinValues(paramSpace.zeroVector());
  QUESO::GslVector paramMaxValues(paramSpace.zeroVector());
  paramMinValues[0] = 8.;
  paramMaxValues[0] = 11.;

  QUESO::BoxSubset<QUESO::GslVector,QUESO::GslMatrix> paramDomain("param_",
      paramSpace, paramMinValues, paramMaxValues);

  // SIP Step 3 of 6: Instantiate the likelihood object to be used by QUESO.
  Likelihood<QUESO::GslVector, QUESO::GslMatrix> lhood("like_", paramDomain);

  // SIP Step 4 of 6: Define the prior RV
  QUESO::UniformVectorRV<QUESO::GslVector,QUESO::GslMatrix> priorRv("prior_",
      paramDomain);

  // SIP Step 5 of 6: Instantiate the inverse problem
  QUESO::GenericVectorRV<QUESO::GslVector,QUESO::GslMatrix>
    postRv("post_",  // Extra prefix before the default "rv_" prefix
           paramSpace);

  QUESO::StatisticalInverseProblem<QUESO::GslVector,QUESO::GslMatrix>
    ip("",          // No extra prefix before the default "ip_" prefix
       NULL,
       priorRv,
       lhood,
       postRv);

  // SIP Step 6 of 6: Solve the inverse problem, that is,
  // set the 'pdf' and the 'realizer' of the posterior RV
  QUESO::GslVector paramInitials(paramSpace.zeroVector());
  priorRv.realizer().realization(paramInitials);

  QUESO::GslMatrix proposalCovMatrix(paramSpace.zeroVector());
  proposalCovMatrix(0,0) = std::pow(std::abs(paramInitials[0]) / 20.0, 2.0);

  ip.solveWithBayesMetropolisHastings(NULL, paramInitials, &proposalCovMatrix);

  // Statistical forward problem (SFP): find the max distance
  // traveled by an object in projectile motion; input pdf for 'g'
  // is the solution of the SIP above.

  // SFP Step 1 of 6: Instantiate the parameter *and* qoi spaces.
  // SFP input RV = FIP posterior RV, so SFP parameter space
  // has been already defined.
  QUESO::VectorSpace<QUESO::GslVector,QUESO::GslMatrix> qoiSpace(env, "qoi_",
      1, NULL);

  // SFP Step 2 of 6: Instantiate the parameter domain
  // NOTE: Not necessary because input RV of the SFP = output RV of SIP.
  // Thus, the parameter domain has been already defined.

  // SFP Step 3 of 6: Instantiate the qoi object to be used by QUESO.
  Qoi<QUESO::GslVector, QUESO::GslMatrix, QUESO::GslVector, QUESO::GslMatrix>
    qoi("qoi_", paramDomain, qoiSpace);

  // SFP Step 4 of 6: Define the input RV
  // NOTE: Not necessary because input RV of SFP = output RV of SIP (postRv).

  // SFP Step 5 of 6: Instantiate the forward problem
  QUESO::GenericVectorRV<QUESO::GslVector, QUESO::GslMatrix> qoiRv("qoi_",
      qoiSpace);

  QUESO::StatisticalForwardProblem<QUESO::GslVector, QUESO::GslMatrix,
    QUESO::GslVector, QUESO::GslMatrix> fp("", NULL, postRv, qoi, qoiRv);

  // SFP Step 6 of 6: Solve the forward problem
  fp.solveWithMonteCarlo(NULL);
}
\end{lstlisting}

\begin{lstlisting}[caption=File \texttt{gravity\_likelihood.h}.,
                 label=code:gravity_like_h]
template<class V, class M>
class Likelihood : public QUESO::BaseScalarFunction<V, M>
{
public:
  Likelihood(const char * prefix, const QUESO::VectorSet<V, M> & domain);
  virtual ~Likelihood();
  virtual double lnValue(const V & domainVector, const V * domainDirection,
      V * gradVector, M * hessianMatrix, V * hessianEffect) const;
  virtual double actualValue(const V & domainVector, const V * domainDirection,
      V * gradVector, M * hessianMatrix, V * hessianEffect) const;

private:
  std::vector<double> m_heights; // heights
  std::vector<double> m_times;   // times
  std::vector<double> m_stdDevs; // uncertainties in time measurements
};
\end{lstlisting}

\begin{lstlisting}[caption=File \texttt{gravity\_likelihood.C}.,
                 label=code:gravity_like_C]
#include <gravity_likelihood.h>

template<class V, class M>
Likelihood<V, M>::Likelihood(const char * prefix,
    const QUESO::VectorSet<V, M> & domain)
  : QUESO::BaseScalarFunction<V, M>(prefix, domain),
    m_heights(0),
    m_times(0),
    m_stdDevs(0)
{
  // Observational data
  double const heights[] = {10, 20, 30, 40, 50, 60, 70, 80, 90, 100, 110,
                            120, 130, 140};

  double const times  [] = {1.41, 2.14, 2.49, 2.87, 3.22, 3.49, 3.81, 4.07,
                            4.32, 4.47, 4.75, 4.99, 5.16, 5.26};

  double const stdDevs[] = {0.020, 0.120, 0.020, 0.010, 0.030, 0.010, 0.030,
                            0.030, 0.030, 0.050, 0.010, 0.040, 0.010, 0.09};

  std::size_t const n = sizeof(heights) / sizeof(*heights);
  m_heights.assign(heights, heights + n);
  m_times.assign(times, times + n);
  m_stdDevs.assign(stdDevs, stdDevs + n);
}

template<class V, class M>
Likelihood<V, M>::~Likelihood()
{
  // Deconstruct here
}

template<class V, class M>
double
Likelihood<V, M>::lnValue(const V & domainVector, const V * domainDirection,
    V * gradVector, M * hessianMatrix, V * hessianEffect) const
{
  double g = domainVector[0];

  double misfitValue = 0.0;
  for (unsigned int i = 0; i < m_heights.size(); ++i) {
    double modelTime = std::sqrt(2.0 * m_heights[i] / g);
    double ratio = (modelTime - m_times[i]) / m_stdDevs[i];
    misfitValue += ratio * ratio;
  }

  return -0.5 * misfitValue;
}

template<class V, class M>
double
Likelihood<V, M>::actualValue(const V & domainVector,
    const V * domainDirection, V * gradVector, M * hessianMatrix,
    V * hessianEffect) const
{
  return std::exp(this->lnValue(domainVector, domainDirection, gradVector,
        hessianMatrix, hessianEffect));
}

template class Likelihood<QUESO::GslVector, QUESO::GslMatrix>;
\end{lstlisting}

\subsection{Running the gravity example with several processors}

\Queso\ requires \MPI, so any compilation of the user's statistical application
will look like this:
\begin{verbatim}
mpicxx -I/path/to/boost/include -I/path/to/gsl/include \
       -I/path/to/queso/include -L/path/to/queso/lib \
       YOURAPP.C -o YOURAPP -lqueso
\end{verbatim}
This will produce a file in the current directory called \texttt{YOURAPP}.  To
run this application with \Queso\ in parallel, you can use the standard
\texttt{mpirun} command:
\begin{verbatim}
mpirun -np N ./YOURAPP
\end{verbatim}
Here \texttt{N} is the number of processes you would like to give to \Queso.
They will be divided equally amongst the number of chains requested (see
\texttt{env\_numSubEnvironments} below.  If the number of requested chains does
not divide the number of processes, an error is thrown.

Even though the application described in Section \ref{sec:gravity_code} is a
serial code, it is possible to run it using more than one processor, i.e.,
produce multiple chains.  Supposing the user's workstation has $N_p=8$
processors, then, the user my choose to have $N_s = 1, \ldots, 8$
subenvironments. This complies with the requirement that the total number of
processors in the environment (eight) must be a multiple of the specified
number of subenvironments (one).  Each subenvironment has only one processor
because the forward code is serial.

Thus, to build and run the application code with $N_p = 8$, and $N_s=8$
subenvironments, the must set the variable \texttt{env\_numSubEnvironments = 8}
in the input file and enter the following command lines:
\begin{verbatim}
mpirun -np 8 ./gravity_gsl gravity_inv_fwd.inp
\end{verbatim}
The steps above will create a total number of 8 raw chains, of size defined by
the variable \texttt{ip\_mh\_rawChain\_size}. \Queso\ internally combines these
8 chains into a single chain of size $8\;
\times\,$\texttt{ip\_mh\_rawChain\_size} and saves it in a file named according
to the variable \texttt{ip\_mh\_rawChain\_dataOutputFileName}.  \Queso\ also
provides the user with the option of writing each chain---handled by its
corresponding processor---in a separate file, which is accomplished by setting
the variable \texttt{ip\_mh\_rawChain\_dataOutputAllowedSet = 0 1 ... Ns-1}.\\

\noindent {\bf Note:} Although the discussion in the previous paragraph refers
to the raw chain of a SIP, the analogous is true for the filtered chains (SIP),
and for the samples employed in the SFP (\texttt{ip\_mh\_filteredChain\_size},
\texttt{fp\_mc\_qseq\_size} and \texttt{fp\_mc\_qseq\_size}, respectively).
See the \Queso\ user's manual for further details.

\subsection{Data post-processing and visualization}
\label{sec:gravity-results}

\subsubsection{Statistical Inverse Problem}

\Queso\ supports both python and Matlab for post-processing. This section
illustrates several forms of visualizing \Queso\ output, and discusses the
results computed by \Queso\ with the code of Section \ref{sec:gravity_code}.
For Matlab-ready commands for post-processing the data generated by \Queso,
refer to the \Queso\ users's manual.

\begin{figure}[hp]
\centering
\subfloat[Raw chain]{\includegraphics[scale=0.35]{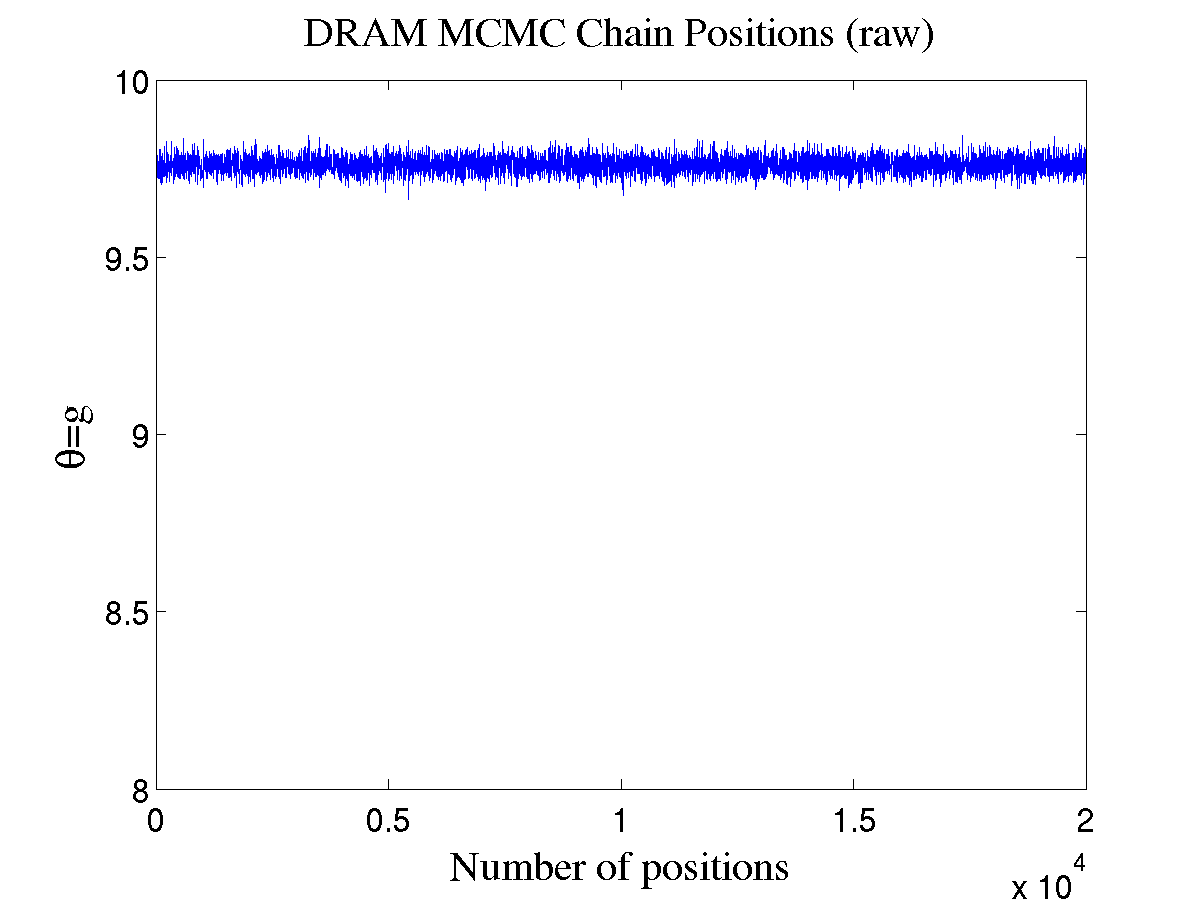}\label{fig:sip_gravity_chain_pos_raw}}
\subfloat[Filtered chain]{\includegraphics[scale=0.35]{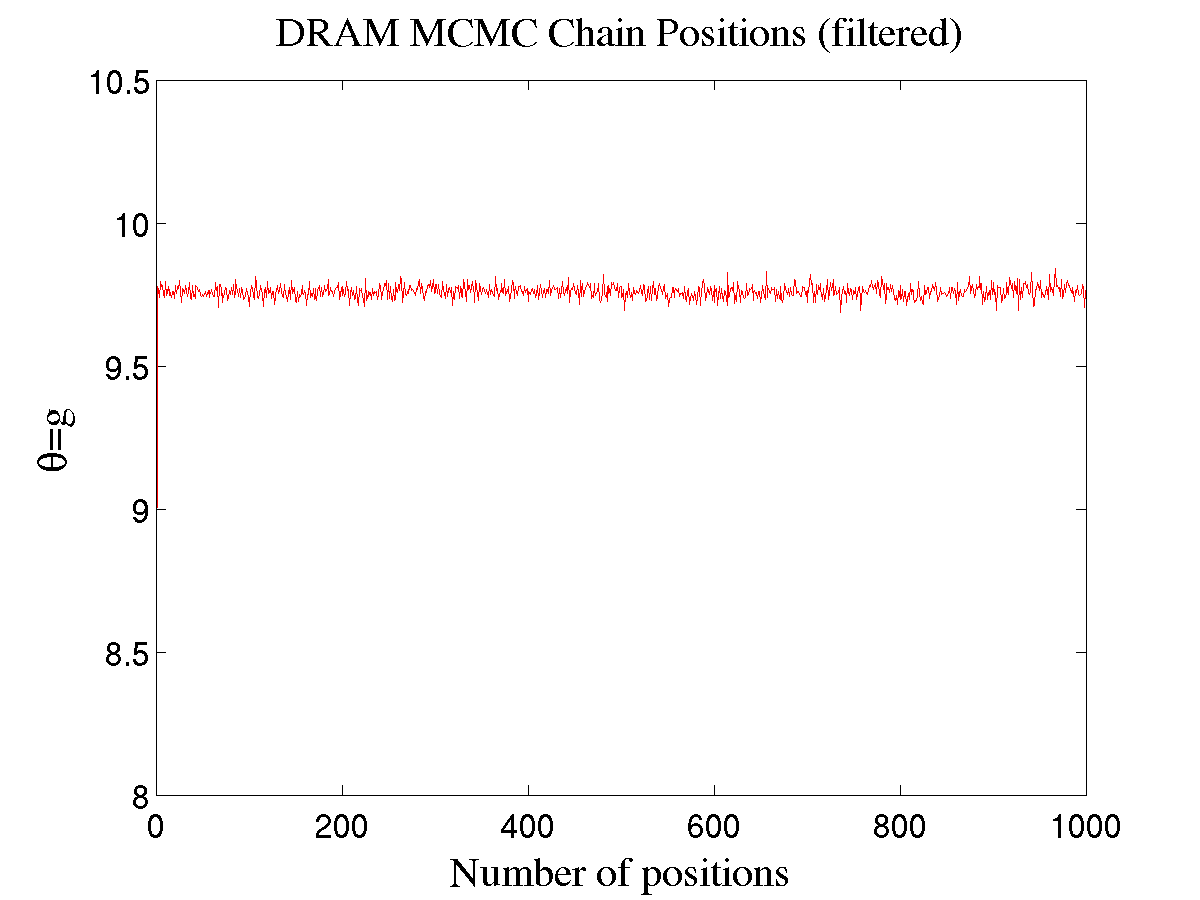}\label{fig:sip_gravity_chain_pos_filtered}}
\vspace*{-10pt}
\caption{MCMC raw chain with \chainsizeresults{} positions and a filtered chain with lag of 20 positions.}
\end{figure}

It is quite simple to plot, using Matlab, the chain of positions used in the
DRAM algorithm implemented within \Queso.
Figure~\ref{fig:sip_gravity_chain_pos_raw} and
Figure~\ref{fig:sip_gravity_chain_pos_filtered} show what raw and filtered
chain output look like, respectively.

\begin{figure}[hp]
\centering
\subfloat[Raw chain]{\includegraphics[scale=0.35]{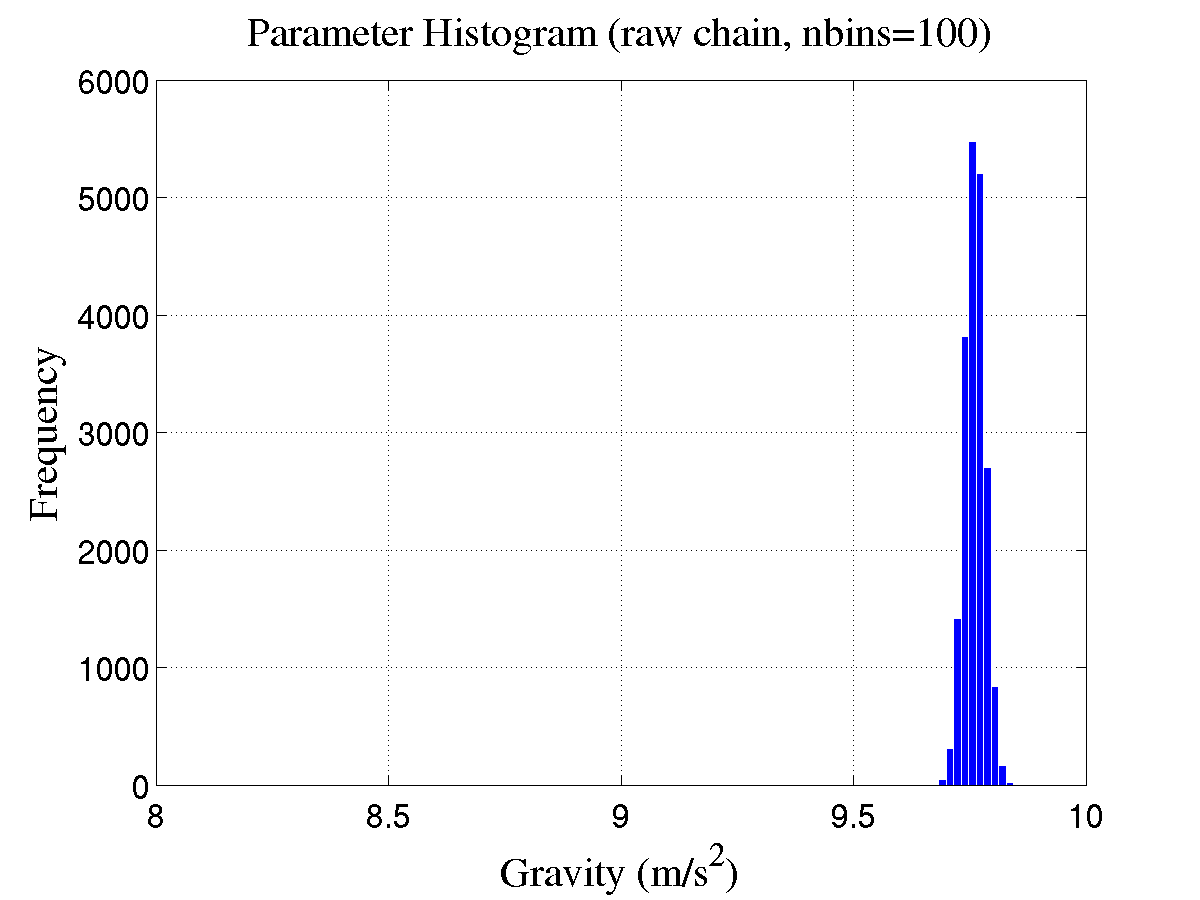}\label{fig:sip_gravity_hist_raw}}
\subfloat[Filtered chain]{\includegraphics[scale=0.35]{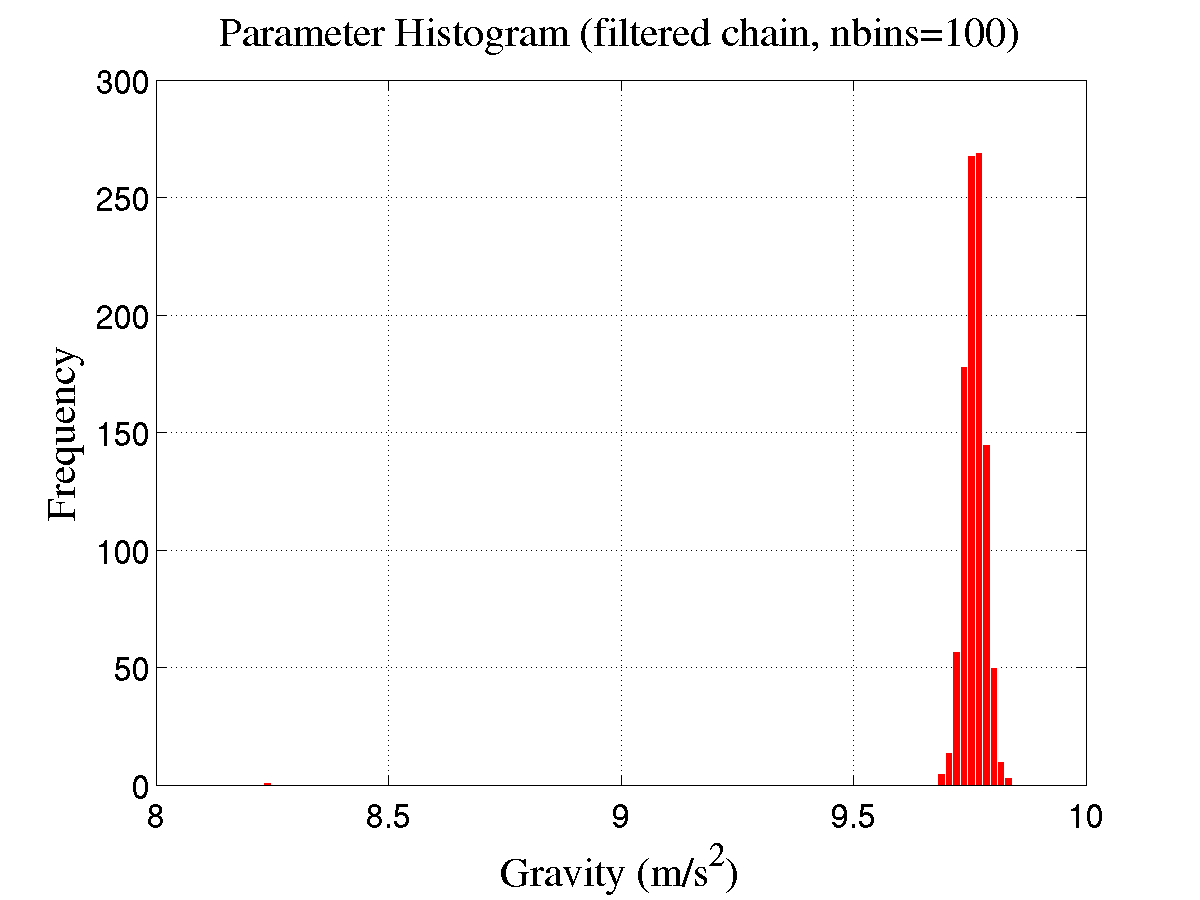}\label{fig:sip_gravity_hist_filtered}}
\vspace*{-10pt}
\caption{Histograms of parameter $\theta=g$. }
\end{figure}

Predefined Matlab and numpy/matplotlib functions exist for converting the raw
or filtered chains into histograms.  The resulting output can be seen in
Figure~\ref{fig:sip_gravity_hist_raw} and
Figure~\ref{fig:sip_gravity_hist_filtered} respectively.

\begin{figure}[hp]
\centering
\subfloat[Raw chain]{\includegraphics[scale=0.35]{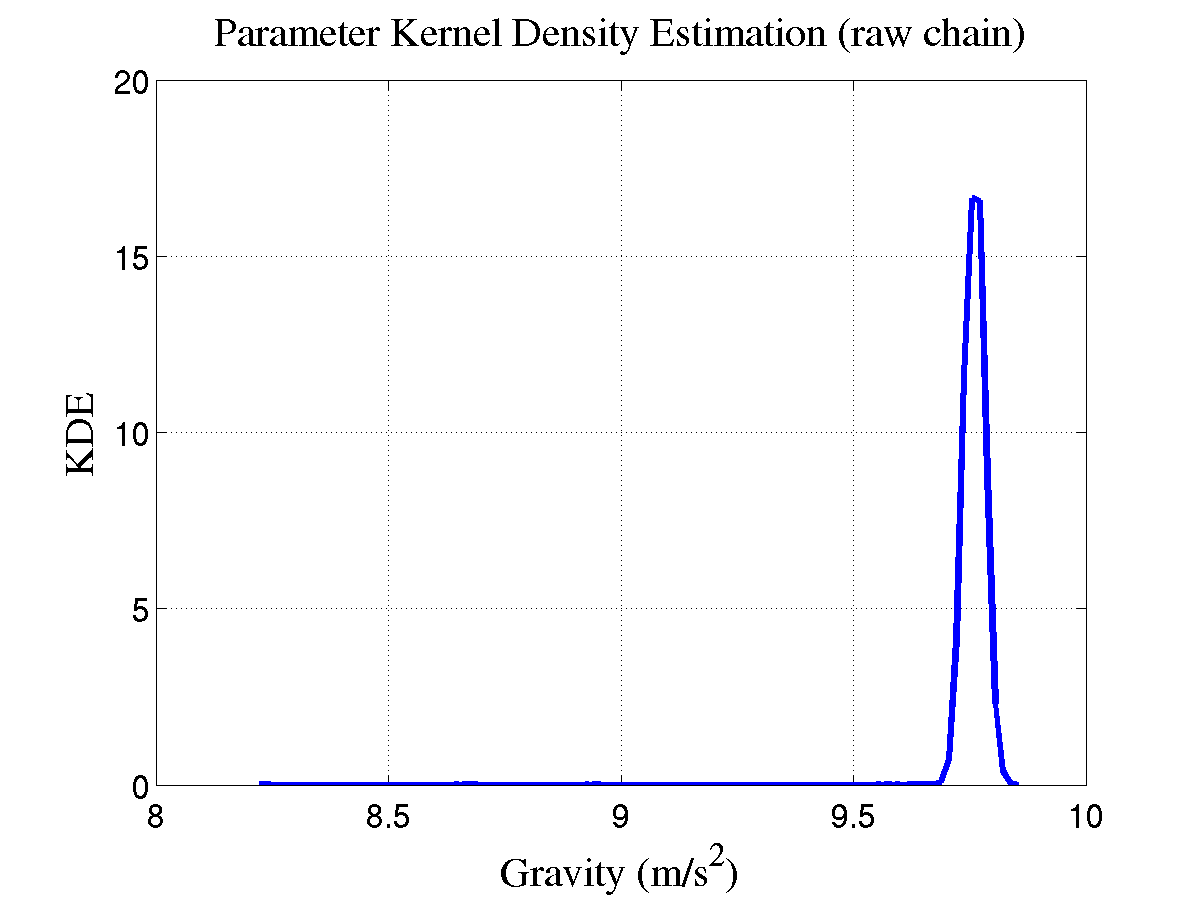}\label{fig:sip_gravity_kde_raw}}
\subfloat[Filtered chain]{\includegraphics[scale=0.35]{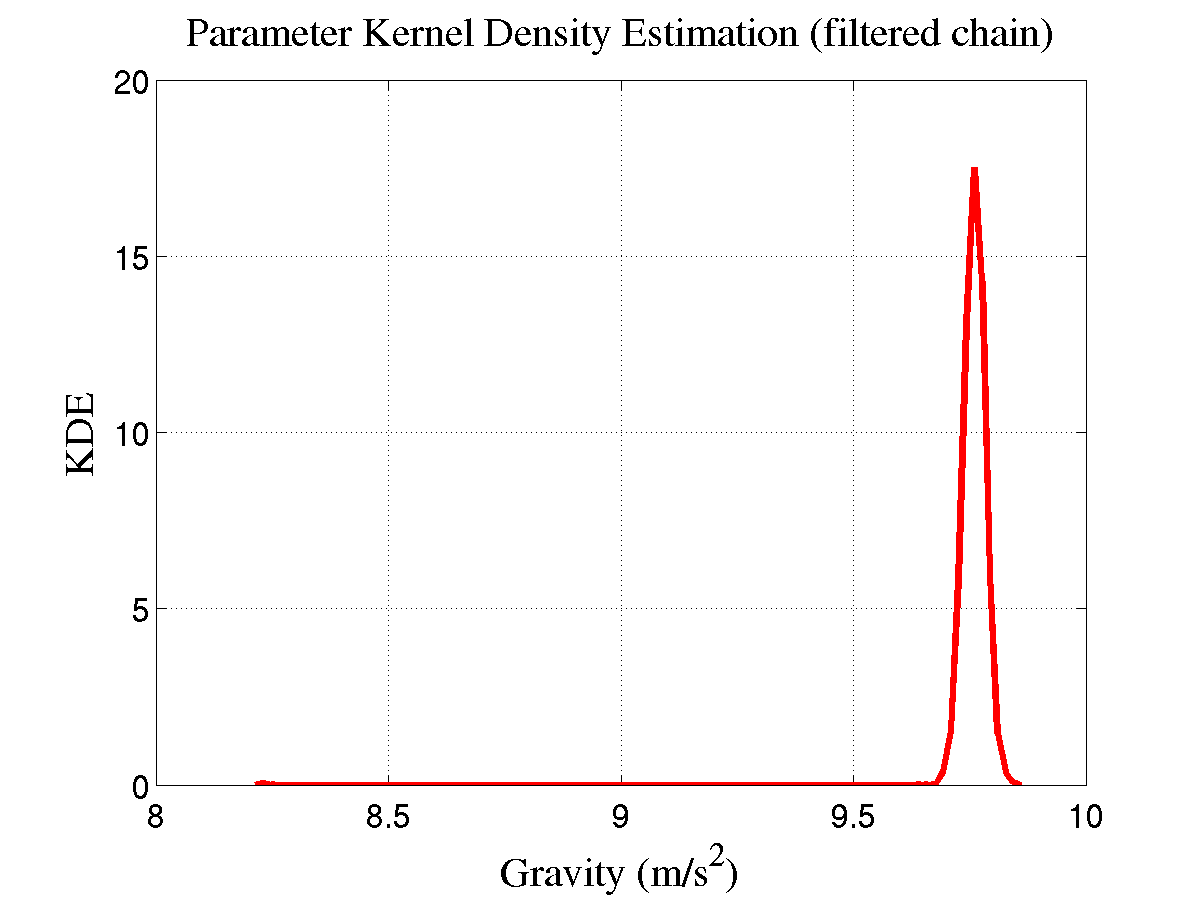}\label{fig:sip_gravity_kde_filtered}}
\vspace*{-10pt}
\caption{Kernel Density Estimation. }
\end{figure}

There are also standard built-in functions in Matlab and SciPy to compute
kernel density estimates.  Resulting output for the raw and filtered chains can
be seen in Figure~\ref{fig:sip_gravity_kde_raw} and
Figure~\ref{fig:sip_gravity_kde_filtered}, respectively.

\subsection{Infinite-dimensional inverse problems}
\label{sec:infd}

\Queso\ has functional but limited support for solving infinite dimensional
inverse problems.  Infinite-dimensional inverse problems are problems for which
the posterior distribution is formally defined on a function space.  After
implementation, this distribution will lie on a discrete space, but the MCMC
algorithm used is robust to mesh refinement of the underlying function space.

There is still substantial work to be done to bring the formulation of these
class of inverse problems in \Queso\ in line with that of the
finite-dimensional counterpart described above, but what currently exists in
\Queso\ is usable.  The reason for the departure in design pattern to that of
the finite-dimensional code is that for infinite-dimensional problems, \Queso\
must be agnostic to any underlying vector type representing the random
functions that are sampled.  To achieve this, a finite element backend is
needed to represent functions.  There are many choices of finite element
libraries that are freely available to download and use, and the design of the
infinite dimensional part of \Queso\ is such that addition of new backends
should be attainable without too much effort.  \Libmesh\ is the default and
only choice currently available in \Queso.  \Libmesh\ is open source and freely
available to download and use.  Visit the \Libmesh\ website for further
details:  \url{http://libmesh.github.io}

We proceed with showing a concrete example of how to formulate an infinite
dimensional inverse problem in \Queso.

First, we assume the user has access to a \texttt{libMesh::Mesh} object on
which their forward problem is defined.  In what follows, we shall call this
object \texttt{mesh}.

\subsubsection{Defining the prior}

Currently, the only measure you can define is a Gaussian measure.  This is
because Gaussian measures are well-defined objects on function space and their
properties are well understood.

To define a Gaussian measure on function space, one needs a mean function and
a covariance operator.  \Queso\ has a helper object to help the user build
functions and operators called \texttt{FunctionOperatorBuilder}.  This object
has properties that are set by the user that define the type and order of the
finite elements used by \Libmesh\ to represent functions:
\begin{verbatim}
// Use a helper object to define some of the properties of our samples
QUESO::FunctionOperatorBuilder fobuilder;
fobuilder.order = "FIRST";
fobuilder.family = "LAGRANGE";
fobuilder.num_req_eigenpairs = num_pairs;
\end{verbatim}
This object will be passed to the constructors of functions and operators and
will instruct \Libmesh, in this case, to use first-order Lagrange finite
elements.  The \texttt{num\_req\_eigenpairs} variable dictates how many
eigenpairs to solve for in an eigenvalue problem needed for the construction of
random functions.  The more eigenpairs used in the construction of Gaussian
random functions, the more high-frequency information is present in the
function.  The downside to asking for a large number of eigenpairs is that the
solution of the eigenvalue problem will take longer.  Solving the eigenvalue
problem, however, is a one-time cost.  The details of the construction of
Gaussian random fields can be found in
\cite{Cotter2012,Bogachev1998,Lifshits1995}.  To define a function, one can do
the following:
\begin{verbatim}
QUESO::LibMeshFunction mean(fobuilder, mesh);
\end{verbatim}
This function is intialized to be exactly zero everywhere.  For more
fine-grained control over point-values, one can access the internal \Libmesh\
\texttt{EquationSystems} object using the \texttt{get\_equation\_systems()}
method.

Specifying a Gaussian measure on a function space is often more convenient to
do in terms of the precision operator rather than the covariance operator.
Currently, the only precision operators available in \Queso\ are powers
of the Laplacian operator.  However, the design of the class hierarchy for
precision operators in \Queso\ should be such that implementation of other
operators is easily achievable.  To create a Laplacian operator in \Queso\ one
can do the following:
\begin{verbatim}
QUESO::LibMeshNegativeLaplacianOperator precision(fobuilder, mesh);
\end{verbatim}
The Gaussian measure can then be defined by the mean and precision above
(where the precision can be taken to a power) as such:
\begin{verbatim}
QUESO::InfiniteDimensionalGaussian mu(env, mean, precision, alpha, beta);
\end{verbatim}
Here \texttt{beta} is the coefficient of the precision operator, and
\texttt{alpha} is the power to raise the precision operator to.

\subsubsection{Defining the likelihood}

Defining the likelihood is very similar to the ball drop example.  We have to
subclass \texttt{InfiniteDimensional\allowbreak LikelihoodBase} and implement the
\texttt{evaluate(FunctionBase \& flow)} method.  This method should return the
logarithm of the likelihood distribution evaluated at the point \texttt{flow}.

One's specific likelihood implementation will vary from problem to problem, but
an example, which is actually independent of \texttt{flow}, is shown here for
completeness:
\begin{verbatim}
double
Likelihood::evaluate(QUESO::FunctionBase & flow)
{
  const double obs_stddev = this->obs_stddev();
  const double obs = gsl_ran_gaussian(this->r, obs_stddev);
  return obs * obs / (2.0 * obs_stddev * obs_stddev);
}
\end{verbatim}
The reader is reminded that a full working implementation of this example is
available in the source tree.  See \url{http://libqueso.com}.

\subsubsection{Sampling the posterior}

The following code will use the prior and the likelihood defined above to setup
the inverse problem and start sampling:
\begin{verbatim}
QUESO::InfiniteDimensionalMCMCSamplerOptions opts(env, "");

// Set the number of iterations to do
opts.m_num_iters = 1000;

// Set the frequency with which we save samples
opts.m_save_freq = 10;

// Set the RWMH step size
opts.m_rwmh_step = 0.1;

// Construct the sampler, and set the name of the output file (will only
// write HDF5 files)
QUESO::InfiniteDimensionalMCMCSampler s(env, mu, llhd, &opts);

for (unsigned int i = 0; i < opts.m_num_iters; i++) {
  s.step();
  if (i % 100 ==  0) {
    std::cout << "sampler iteration: " << i << std::endl;
    std::cout << "avg acc prob is: " << s.avg_acc_prob() << std::endl;
    std::cout << "l2 norm is: " << s.llhd_val() << std::endl;
  }
}
\end{verbatim}

The infinite dimensional inverse problem work is still considered experimental,
but should produce meaningful results for a large class of simple problems.
Work is ongoing to bring the user interface inline with that of the
finite-dimensional inverse problem API.

\section{Extensibility}
\label{sec:extend}

\Queso\ is written in C++.  The choice of the language inspired design decisions
that the user can take advantage of.  One such benefit of having a well-defined
inverse problem setup and workflow is that the user is offered the freedom to
extend many of the abstract base classes in \Queso.  A good example of this
we have seen already is the specification of the likelihood distribution by
subclassing \texttt{BaseScalarFunction}.

In this section we will take this a step further and show how to extend some of
the other classes in \Queso\ to define a custom prior measure.  All of the
classes we deal with here have their relationships with other objects discussed
in section \ref{sec:internals}.

\subsection{Custom priors}

We will look at one of the existing measures in \Queso\ to get a feel for a how
a measure \Queso\ is built.  Take, for example, the Gamma distribution.

In \Queso, the user will interact with a Gamma measure by instantiating a
\texttt{GammaVectorRV} class.  This object has two main members that \Queso\ is
interested in, an object representing a probability distribution function, and
an object called a `realizer' through with random variates are drawn.  These
classes are called \texttt{GammaJointPdf} and \texttt{GammaVectorRealizer},
respectively.

The user does not, usually, need to interact with the probability distribution
function or the realizer; these are objects that \Queso\ will utilize during
the execution of the Markov chain Monte Carlo procedure.

\subsubsection{PDF objects}

Probability distribution functions are represented by C++ objects.  If the user
wishes to create a custom prior measure, for example, then they will also have
to implement a probability distribution class.  The probability distribution
class must derive from the \texttt{BaseJointPdf}.  The \texttt{BaseJoinPdf}
class subclasses from \texttt{BaseScalarFunction}, as we have seen before, and
therefore any probability distribution class must implement the \texttt{lnValue}
and \texttt{actualValue} methods.  These methods have exactly the same purpose
as when the user defines their likelihood.  That is, \texttt{lnValue} returns
the log of the probability distribution function evaluated at
\texttt{domainVector} and \texttt{actualValue} returns the actual value of the
distribution evaluated at \texttt{domainVector}.

\texttt{BaseJointPdf} has an extra method called
\texttt{computeLogOfNomalizationFactor} and so this must also be implemented.
This method computes the logarithm of the normalizing constant of the
probability distribution.  If it is known analytically, the user can implement
it here.  For many distributions, this is not known analytically.  In these
circumstances one can use the \texttt{numSamples} argument to approximate this
quantity using samples from the distribution instead.  A basic algorithm for
computing the log of the normalizing constant from samples is implemented in
the \texttt{commonComputeLogOfNormalizationFactor} method of
\texttt{BaseJointPdf}.  Indeed the computation of the log of the normalization
constant for the Gamma distribution is handed off to this method:
\begin{verbatim}
template<class V, class M>
double
GammaJointPdf<V, M>::computeLogOfNormalizationFactor(
    unsigned int numSamples,
    bool updateFactorInternally) const
{
  value =
    BaseJointPdf<V,M>::commonComputeLogOfNormalizationFactor(
      numSamples, updateFactorInternally);
  return value;
}
\end{verbatim}

Notice that when we defined a custom likelihood object we only subclassed
\texttt{BaseScalarFunction} and not \texttt{BaseJointPdf}.  This is because for
most applications the likelihood is not a probability distribution since it
does not integrate to 1.  Furthermore, it avoids needing to implement the
\texttt{computeLogOfNormalizationFactor} method.  This is because the
normalizing constant is usually not known analytically and computing it by
samples is often intractable for large engineering problems.  Note, however,
that the normalizing constant for the likelihood is not needed since MCMC
methods do not require knowledge of any normalizing constant in order to draw
random samples.  This is crystallized in the following section.

\subsubsection{Realizer objects}

Realizer objects are objects that \Queso\ interacts with to draw random samples
from the appropriate distribution.  A realizer object must subclass
\texttt{BaseVectorRealizer} and must therefore implement the
\texttt{realization(V \& nextValues) const} method.  This method fills the
\texttt{nextValues} vector with a random draw from the associated distribution.
The size of the vector \texttt{nextValues} is equal to the dimension of the
state space on which the measure is defined.

In the case of the Gamma distribution, \Queso\ falls back to \Gsl\ to draw
samples that are Gamma distributed.

A warning to the user: it is possible to define a measure on a space that is improper.  
In this case drawing realizations from the associated realizer object 
produces meaningless results.

\subsubsection{Random variable objects}

Random variable objects, named \texttt{*VectorRV} in \Queso, are encapsulating
objects that hold references to the associated probability distribution
function object and the associated realizer object.  A random variable object
must subclass \texttt{BaseVectorRV} which implements the getter methods
\texttt{realizer()} and \texttt{pdf()} that return references to the
realizer object and PDF object, respectively.

The user never has to deal with constructing the PDF object or the realizer
object explicitly.  Construction of these objects is handled by the random
variable object's constructor.

%
%

\section{The QUESO Design and Implementation}
\label{sec:design}

\subsection{Software Engineering}

High quality software is essential for developing, analyzing and scaling up new
UQ algorithmic ideas involving complex simulation codes running on HPC
platforms.  \Queso\ helps researchers to bootstrap statistical inverse problems
for large-scale models widely seen in the physics and engineering domains in
parallel compute environments.  With ongoing effort to enhance the API in terms
of extensibility (see section \ref{sec:future-api}), in the future it will be
possible to quickly prototype new algorithms in a sophisticated computation
environment, rather than first coding and testing them with a scripting
language and only then recoding in a C++/MPI environment.  \Queso\ also allows
researchers to more naturally translate the mathematical language present in
algorithms to a concrete program in the library, and to concentrate their
efforts on algorithmic, load balancing and parallel scalability issues.

We utilize various community tools to manage the \Queso\ development cycle.
Source code traceability is provided via git and the GNU autotools suite is
used to provide a portable, flexible build system, with the standard GNU
package pattern: \texttt{configure; make; make check; make install} steps.  We
also utilize GitHub for project management, which provides a web-based
mechanism to manage releases, milestone developments, issues, bugs, and source
code changes.  In case the build system or application development processes
change, please consult the website (\url{http://libqueso.com}) for a detailed
and up-to-date guide on how to build and install \Queso.

As of the latest \Queso\ release, 0.53.0, the library is comprised of
approximately 73,000 source lines of code,
with the vast majority of this instantiated across approximately 200 C/C++
source files and headers.  At a minimum, \Queso\ compilation requires \MPI\ and
linkage against two external libraries: \Boost\ and \Gsl.  \Queso\ also has
several optional dependencies that enable additional functionality: \Teuchos,
\Grvy, \HDF. The optional infinite dimensional capabilities of \Queso\ in
particular require \Libmesh\ and \HDF.

We employ an active regression testing, with approximately thirty regression
tests, and can test latest GitHub builds using Travis-CI in order to have a
continuous integration analysis of source code commits.

Contributing \Queso\ has been made easy with the recent explosion in popularity
of GitHub.  We employ the feature branch model by Driessen
(\url{http://nvie.com/posts/a-successful-git-branching-model}) and further
instructions for contribution to \Queso\ can be found by mirroring some of the
other contributions we have merged
(\url{https://github.com/libqueso/queso/issues}).


\subsection{\Queso\ internals}
\label{sec:internals}

In this subsection, we show and discuss several of the inheritance diagrams
behind the principle objects in the \Queso\ library. This is in order to:

\begin{itemize}
  \item Document the \Queso\ internal structure
  \item Provide context for leveraging the existing \Queso\ objects in
    extending the library (as in section \ref{sec:extend}).
\end{itemize}

This subsection addresses some of the C++ objects for the finite-dimensional
Bayesian inverse problem.  Objects associated the infinite-dimensional problem
exist, and are available on the online documentation, but are not discussed
here since development work to get the finite- and infinite-dimensinoal APIs
consistent with each other is ongoing.

\texttt{BaseScalarFunction} is a templated base class for handling generic
scalar functions. This provides a high level interface and member functions for
the \Queso\ generic class, \texttt{BaseJointPDF}, which is discussed below.

\texttt{BaseJointPdf} is a templated (base) class for handling joint PDFs.
\begin{figure}[hp]
  \centering
  \includegraphics[scale=0.35]{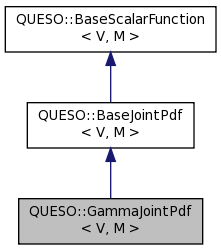}
  \caption{Class Reference for the Gamma Joint PDF.}
  \label{fig:gamma}
\end{figure}
For example, figure \ref{fig:gamma} shows the inheritance of the Gamma joint
PDF class, which is a derived class from the \texttt{BaseScalarFunction} class.
\Queso\ presently has several provided joint PDFs for a wide variety of
statistical distributions, including: \texttt{InvLogitGaussianJointPdf},
\texttt{ConcatenatedJointPdf}, \texttt{GaussianJointPdf},
\texttt{BaseJointPdf}, \texttt{BayesianJointPdf}, \texttt{LogNormalJointPdf},
\texttt{PoweredJointPdf}, \texttt{BetaJointPdf}, \texttt{GammaJointPdf},
\texttt{InverseGammaJointPdf}, \texttt{WignerJointPdf},
\texttt{GenericJointPdf}, \texttt{UniformJointPdf}, \texttt{JeffreysJointPdf},
\texttt{GenericScalarFunction}, and \texttt{ConstantScalarFunction}. However,
implementing a new distribution is intended to be straightforward, and is
detailed in section \ref{sec:extend}.

\begin{figure}[hp]
\centering
\includegraphics[scale=0.35]{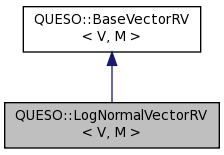}
\caption{Class Reference for the LogNormal Vector RV.}
\label{fig:lognormal}
\end{figure}

Another useful internal \Queso\ object, \texttt{BaseVectorRV}, is a templated
base class for handling vector random variables.  For example, figure~\ref{fig:lognormal}
 shows the inheritance diagram of the \texttt{LogNormalRV}
class, which is a class that contains member functions and associated utilities
to provide a random vector of draws from a \texttt{LogNormal} distribution.

Presently included in \Queso\ are the following: \texttt{GaussianVectorRV},
\texttt{GenericVectorRV}, \texttt{BetaVectorRV}, \texttt{GammaVectorRV},
\texttt{InverseGammaVectorRV}, \texttt{InvLogitGaussianVectorRV},
\texttt{JeffreysVectorRV}, \texttt{LogNormalVectorRV},
\texttt{UniformVectorRV}, and \texttt{WignerVectorRV}. In other words, nearly
all canonical distributions from classical statistics are already available in
the library. However, as stated above, \Queso\ is designed with extensibility
in mind, and the user can implement any \texttt{*VectorRV} by deriving from the
\texttt{BaseVectorRV} class.  In principle, this permits a series of draws from
any distribution.

Another important base class contained within \Queso\ is the realizer object,
\texttt{BaseVectorRealizer}. A realizer is an object that, simply put, contains
a \texttt{realization()} operation that returns a sample of a random variable.
\texttt{BaseVectorRealizer} is therefore an abstract base class that provides
the necessary interface for sampling from random variables.  As before, the
realizer object contains most of the common statistical distributions. It also
contains a sequence realizer class for storing samples of a MH algorithm.

\section{Algorithms}
\label{sec:algos}

\subsection{DRAM}
\label{sec:algo-dram}

A simple Metropolis-Hastings sampling algorithm \cite{Metropolis1953} can be
improved by adding both ``Delayed Rejection''
\cite{Mira2001,Mira2001a,Tierney1999,Green2001} and ``Adaptive Metropolis''.
Taken together, these form the ``DRAM'' algorithm, which is available in
\Queso. In particular, the \Queso\ implements the DRAM algorithm of Haario,
Laine, Mira and Saksman~\cite{Haario2006}.

A ``vanilla'' Metropolis-Hastings sampler involves a proposal at each step, and
accepts or rejects this proposal based on the ratio between proposal and
prior likelihoods.  Typically, the proposal is drawn from some fixed
distribution, such as a Gaussian distribution with fixed covariance and a mean
centered at the value of the current state of the chain.  However, this has
several deficiencies. Should the proposal variance be set too high, many
proposals will be rejected. This is undesirable, as it increases the
auto-correlation of the chain.  Furthermore, should the target distribution
deviate greatly from the proposal distribution, the proposal will not match the
local shape of the distribution, resulting in poor sampling.

Delayed rejection attempts to circumvents these issues.  Before rejecting a
sample, a series of back-up proposals each with successively smaller jumps
in state space are pushed through the Metropolis-Hasting acceptance probability
rejection.  They are tested in order of decreasing jump size and if one of them
is accepted, the sampler continues.  If they are all rejected, the sampler
rejects the sample and starts again.

Conversely, when the proposal variance is too small to efficiently sample the
target distribution, the sampler will randomly walk through regions of higher
likelihood in the posterior distribution, without efficiently sampling the
tails. This results in too high an acceptance rate.

In order to mitigate this, Adaptive Metropolis sampling continuously adapts the
proposal covariance. This is accomplished by using the covariance of the
history of the Markov chain as the proposal covariance matrix of the Gaussian
proposal distribution instead of the arbitrary proposal covariance imposed at
the start.  Adapting the proposal to match the posterior covariance structure
results in a better chain performance than a static proposal covariance.

\subsection{Multi-level}
\label{sec:algos-ml}

Multi-level Monte Carlo \cite{Cheung2012} is an algorithm available in \Queso\
that attempts to sample probability distributions with multiple modes.
Sampling multi-modal distributions is a heavily researched topic.  The way
Multi-level Monte Carlo attempts to solve the problem of meta-stability in
Markov chains is by `heating up' the posterior distribution to flatten out some
of the modes, allowing a Markov chain to sample the flattened distribution and
then `cooling down' the posterior distribution before doing a final sampling
run.  The idea is identical to that of simulated tempering or simulated
annealing, except that the Multi-level algorithm allows for convenient and
efficient computation of the posterior normalizing constant.  This constant is
usually intractable to compute but is essential for Bayesian model selection
purposes.

\subsection{Pre-conditioned Crank-Nicolson}
\label{sec:algo-pcn}

The pre-conditioned Crank-Nicolson proposal \cite{Cotter2012} is used by
\Queso\ for solving infinite-dimensional Bayesian inverse problems (section
\ref{sec:infd}).  This particular form of proposal is typical for sampling on
formally infinite-dimensional spaces since the Metropolis-Hastings acceptance
probability remains unchanged when the state undergoes mesh refinement, a
popular technique in large-scale engineering models involving the solution of
partial different equations by finite element methods.

\section{Input file}

Here we provide some of the default input file options \Queso\ recognizes.  For
detailed descriptions of the behavior of each option and how they interact with
other options, consult the online \Queso\ documentation.  For example, for the
description of each DRAM option, consult the documentation for the
\texttt{MhOptionsValues} object.  For the description of each
\texttt{FullEnvironment} option, see the documentation for the
\texttt{EnvOptionsValues} object.  The documentation for these is available at
\url{http://libqueso.com}.

\begin{center}
\begin{longtable}{l c  m{7cm}}
\caption{Input file options for a QUESO environment.}\\
\label{tab-env-options}
\footnotesize
Option name                      &  Default & Description \\
\midrule\midrule
\ttfamily env\_help                &     & Produces help message for environment class            \\
\ttfamily env\_numSubEnvironments   &  1  &  Number of subenvironments                \\ 
\ttfamily env\_subDisplayFileName   & \ttfamily"." & Output filename for sub-screen writing     \\ 
\ttfamily env\_subDisplayAllowAll   &  0  & Allows all subenvironments to write to output file \\ 
\ttfamily env\_subDisplayAllowedSet & \ttfamily""  & Subenvironments that will write to output file \\ 
\ttfamily env\_displayVerbosity     &  0  & Sets verbosity				         \\ 
\ttfamily env\_syncVerbosity        &  0  & Sets syncronized verbosity             \\ 
\ttfamily env\_seed                 &  0  & Set seed                             \\ 
\bottomrule
\end{longtable}
\end{center}

\begin{center}
\begin{longtable}{l c  m{7cm}}
\caption{Input file options for a QUESO statistical inverse problem.}\\
\label{tab-sip-options}
\rmfamily Option name                    & \rmfamily Default & \rmfamily Description \\
\midrule\midrule
\ttfamily ip\_help                 &     &  \rmfamily Produces help message for statistical inverse problem   \\
\ttfamily ip\_computeSolution      &  1  &  \rmfamily Computes solution process \\
\ttfamily ip\_dataOutputFileName   & "." &  \rmfamily Name of data output file \\
\ttfamily ip\_dataOutputAllowedSet & ""  &  \rmfamily Subenvironments that will write to data output file  \\
\bottomrule
\end{longtable}
\end{center}

\begin{center}
\begin{longtable}{l c} 
\caption{Input file options for a QUESO DRAM solver.}\\
\label{tab-metropolis-hastings-options}
\rmfamily Option Name                                    & \rmfamily Default Value \\
\midrule\midrule
\ttfamily mh\_dataOutputFileName                       & "."   \\ 
\ttfamily mh\_dataOutputAllowAll                       & 0     \\ 
\ttfamily mh\_initialPositionDataInputFileName         & "."   \\ 
\ttfamily mh\_initialPositionDataInputFileType         & "m"   \\ 
\ttfamily mh\_initialProposalCovMatrixDataInputFileName& "."   \\ 
\ttfamily mh\_initialProposalCovMatrixDataInputFileType& "m"   \\ 
\ttfamily mh\_rawChainDataInputFileName                & "."   \\ 
\ttfamily mh\_rawChainDataInputFileType                & "m"   \\ 
\ttfamily mh\_rawChainSize                             & 100   \\ 
\ttfamily mh\_rawChainGenerateExtra                    &  0    \\ 
\ttfamily mh\_rawChainDisplayPeriod                    & 500   \\ 
\ttfamily mh\_rawChainMeasureRunTimes                  &  1    \\ 
\ttfamily mh\_rawChainDataOutputPeriod                 &  0    \\ 
\ttfamily mh\_rawChainDataOutputFileName               & "."   \\ 
\ttfamily mh\_rawChainDataOutputFileType               & "m"   \\ 
\ttfamily mh\_rawChainDataOutputAllowAll               &  0    \\ 
\ttfamily mh\_filteredChainGenerate                    &  0    \\ 
\ttfamily mh\_filteredChainDiscardedPortion            &  0.   \\ 
\ttfamily mh\_filteredChainLag                         &  1    \\ 
\ttfamily mh\_filteredChainDataOutputFileName          & "."   \\ 
\ttfamily mh\_filteredChainDataOutputFileType          & "m"   \\ 
\ttfamily mh\_filteredChainDataOutputAllowAll          &  0   \\ 
\ttfamily mh\_displayCandidates                        &  0    \\ 
\ttfamily mh\_putOutOfBoundsInChain                    &  1    \\ 
\ttfamily mh\_tkUseLocalHessian                        &  0    \\ 
\ttfamily mh\_tkUseNewtonComponent                     &  1    \\ 
\ttfamily mh\_drMaxNumExtraStages                      &  0    \\ 
\ttfamily mh\_drDuringAmNonAdaptiveInt                 &  1    \\ 
\ttfamily mh\_amKeepInitialMatrix                      &  0    \\ 
\ttfamily mh\_amInitialNonAdaptInterval                &  0    \\ 
\ttfamily mh\_amAdaptInterval                          &  0    \\ 
\ttfamily mh\_amAdaptedMatricesDataOutputPeriod        &  0    \\ 
\ttfamily mh\_amAdaptedMatricesDataOutputFileName      & "."   \\ 
\ttfamily mh\_amAdaptedMatricesDataOutputFileType      & "m"   \\ 
\ttfamily mh\_amAdaptedMatricesDataOutputAllowAll      &  0    \\ 
\ttfamily mh\_amEta                                    & 1.    \\ 
\ttfamily mh\_amEpsilon                                & $1\times 10^{-5}$   \\ 
\ttfamily mh\_enableBrooksGelmanConvMonitor            & 0    \\ 
\ttfamily mh\_BrooksGelmanLag                          & 100   \\ 
\bottomrule
\end{longtable}
\end{center}

\begin{center}
\begin{longtable}{l c} 
\caption{Input file options for a QUESO Multilevel solver.}\\
\label{tab-Multilevel-options}
\rmfamily Option Name                                    & \rmfamily Default Value \\
\midrule\midrule
\ttfamily ml\_restartOutput\_levelPeriod       & 0    \\ 
\ttfamily ml\_restartOutput\_baseNameForFiles  & "."  \\ 
\ttfamily ml\_restartOutput\_fileType          & "m"  \\ 
\ttfamily ml\_restartInput\_baseNameForFiles   & "."  \\ 
\ttfamily ml\_restartInput\_fileType           & "m"  \\ 
\ttfamily ml\_stopAtEnd                                 & 0    \\
\ttfamily ml\_dataOutputFileName                        & "."  \\
\ttfamily ml\_dataOutputAllowAll                        & 0    \\
\ttfamily ml\_loadBalanceAlgorithmId                    & 2    \\
\ttfamily ml\_loadBalanceTreshold                       & 1.0  \\
\ttfamily ml\_minEffectiveSizeRatio                     & 0.85 \\
\ttfamily ml\_maxEffectiveSizeRatio                     & 0.91 \\
\ttfamily ml\_scaleCovMatrix                            & 1    \\
\ttfamily ml\_minRejectionRate                          & 0.50 \\
\ttfamily ml\_maxRejectionRate                          & 0.75 \\
\ttfamily ml\_covRejectionRate                          & 0.25 \\
\ttfamily ml\_minAcceptableEta                          & 0.   \\
\ttfamily ml\_totallyMute                               & 1    \\
\ttfamily ml\_initialPositionDataInputFileName          & "."  \\
\ttfamily ml\_initialPositionDataInputFileType          & "m"  \\
\ttfamily ml\_initialProposalCovMatrixDataInputFileName & "."  \\
\ttfamily ml\_initialProposalCovMatrixDataInputFileType & "m"  \\
\ttfamily ml\_rawChainDataInputFileName                 & "."  \\
\ttfamily ml\_rawChainDataInputFileType                 & "m"  \\
\ttfamily ml\_rawChainSize                              & 100  \\
\ttfamily ml\_rawChainGenerateExtra                     & 0    \\
\ttfamily ml\_rawChainDisplayPeriod                     & 500  \\
\ttfamily ml\_rawChainMeasureRunTimes                   & 1    \\
\ttfamily ml\_rawChainDataOutputPeriod                  & 0    \\
\ttfamily ml\_rawChainDataOutputFileName                & "."  \\
\ttfamily ml\_rawChainDataOutputFileType                & "m"  \\
\ttfamily ml\_rawChainDataOutputAllowAll                & 0    \\
\ttfamily ml\_filteredChainGenerate                     & 0    \\
\ttfamily ml\_filteredChainDiscardedPortion             & 0.   \\
\ttfamily ml\_filteredChainLag                          & 1    \\
\ttfamily ml\_filteredChainDataOutputFileName           & "."  \\
\ttfamily ml\_filteredChainDataOutputFileType           & "m"  \\
\ttfamily ml\_filteredChainDataOutputAllowAll           & 0    \\
\ttfamily ml\_displayCandidates                         & 0    \\
\ttfamily ml\_putOutOfBoundsInChain                     & 1    \\
\ttfamily ml\_tkUseLocalHessian                         & 0    \\
\ttfamily ml\_tkUseNewtonComponent                      & 1    \\
\ttfamily ml\_drMaxNumExtraStages                       & 0    \\
\ttfamily ml\_drScalesForExtraStages                    & 0    \\
\ttfamily ml\_drDuringAmNonAdaptiveInt                  & 1    \\
\ttfamily ml\_amKeepInitialMatrix                       & 0    \\
\ttfamily ml\_amInitialNonAdaptInterval                 & 0    \\
\ttfamily ml\_amAdaptInterval                           & 0    \\
\ttfamily ml\_amAdaptedMatricesDataOutputPeriod         & 0    \\
\ttfamily ml\_amAdaptedMatricesDataOutputFileName       & "."  \\ 
\ttfamily ml\_amAdaptedMatricesDataOutputFileType       & "m"  \\ 
\ttfamily ml\_amAdaptedMatricesDataOutputAllowAll       & 0    \\ 
\ttfamily ml\_amEta                                     & 1.   \\ 
\ttfamily ml\_amEpsilon                                 & 1.e-5\\ 
\bottomrule
\end{longtable}
\end{center}

\begin{center}
\begin{longtable}{l c} 
\caption{Input file options for a QUESO pCN solver.}\\
\label{tab-pcn-options}
\rmfamily Option Name                                    & \rmfamily Default Value \\
\midrule\midrule
\ttfamily infmcmc\_dataOutputDirName & "chain"   \\
\ttfamily infmcmc\_dataOutpuFileName & "out.h5"  \\
\ttfamily infmcmc\_num\_iters        & 1000      \\
\ttfamily infmcmc\_save\_freq        & 1         \\
\ttfamily infmcmc\_rwmh\_step        & 1e-2      \\
\bottomrule
\end{longtable}
\end{center}

\section{Conclusions}

We conclude this chapter with a discussion of several of the areas the \Queso\
development team are investing time into implementing, extending and improving
along with some of the newest features recently made available in v0.53.0.
Previously, we have covered only the basics of how to interact with \Queso\ and
to provide a resource that is accessible and can be used to bootstrap a user's
statistical inverse problem quickly and efficiently.  With this in mind, there
are still many areas in which \Queso\ can improve to become more user-friendly,
consistent, and extensible.  In what follows, we discuss some major areas of
development that would likely encourage wide-spread adoption of \Queso\ in the
computational applied mathematics and engineering community.

\subsection{\Queso-provided likelihoods}

In many large-scale physics and engineering-based experimental settings, it is
often the case that observations of a physical quantity are performed several
times.  These observations are then averaged to homogenize the effect of
experimental observation error.  In the case of independent experimental
errors, this average will be normally distributed.  Therefore, a reasonable
choice for a likelihood in many applications would be a Gaussian.

At present, the user must derive from \texttt{BaseScalarFunction} and implement
\texttt{lnValue} explicitly.  This is a tedious task if all is needed is the
standard Gaussian error in the Euclidean 2-norm,
\begin{equation}
  \mathbb{P}(y | \theta) = Z \exp \left( \frac12
  \left( \mathcal{G}(\theta) - y \right)^{\top} \Sigma^{-1}
  \left( \mathcal{G}(\theta) - y \right) \right),
  \label{eqn:likelihood}
\end{equation}
where $Z$ is a normalizing constant.

A recently-released and much leaner approach is to provide an abstract base
class of \texttt{BaseScalarFunction} called \texttt{BaseGaussianLikelihood}
with a pure virtual method called \texttt{evaluateModel} that asks for the
output of the map $\mathcal{G}$ at the point \texttt{domainVector}.  Equipped
with an implementation of \texttt{lnValue} that computes the log of
\eqref{eqn:likelihood}, the user would only need to provide $\Sigma$ and $y$,
which can be passed in from the constructor.  An example follows:
\begin{verbatim}
template<class V, class M>
class Likelihood : public
    QUESO::GaussianLikelihoodScalarCovariance<V, M>
{
public:

  Likelihood(const char * prefix,
      const QUESO::VectorSet<V, M> & domain,
      const V & observations, double variance)
    : QUESO::GaussianLikelihoodScalarCovariance<V, M>(
        prefix, domain,
        observations, variance)
  { }

  virtual ~Likelihood() { }

  virtual void evaluateModel(const V & domainVector,
      const V * domainDirection,
      V & modelOutput, V * gradVector,
      M * hessianMatrix,
      V * hessianEffect) const
  {
    // Evaluate model and fill up the modelOutput
    // variable
    int dim = modelOutput.sizeLocal();
    for (unsigned int i = 0; i < dim; i++) {
      modelOutput[i] = 1.0;  // Replace this with
                             // the output from
                             // your model
    }
  }
};
\end{verbatim}
Here the user would pass an instance of \texttt{Likelihood} to
\texttt{StatisticalInverseProblem}, as per usual.

Extensions of this idea are also available, where one wishes to treat $\Sigma$
as a hyper-parameter to be sampled along with $\theta$ in so-called
`hierarchical Bayesian' methods.  The design described above is easily applied
to this situation.

Ongoing work is being invested in developing other pre-made likelihood objects
representing other likelihood forms that are commonly used.

\subsection{Emulators}

The two main forms of emulation used in the statistical modeling community are
Gaussian processes and generalized polynomial chaos.  These are both important
methods in statistical inference as they can considerably reduce the
computational cost of computing the posterior.

Gaussian process emulators, similar to the ready-made Gaussian likelihoods
discussed in the previous section, are also a form of baked likelihood, but
where the user is not required to implement a method returning the output of
$\mathcal{G}$.  For Gaussian process emulators the user would only need to
instantiate an emulator with a specific dataset and observational error
covariance matrix.  The rest of the statistical application the user writes is
identical to any other statistical application and the output (samples) is
processed as per usual.

Generalized polynomial chaos methods require different algorithms for
solution, since no Markov chain Monte Carlo is done.  This type of emulator is
not currently on the \Queso\ development roadmap for the near future, but
contributions in the area are more than welcome.

As of \Queso\ v0.53.0, the only supported emulator is a linear interpolation of
model output values.  Interested users should consult the documentation and,
in particular, the example called \texttt{4d\_interp.C}.

\subsection{API considerations}
\label{sec:future-api}

As mentioned in the infinite-dimensional example, the infinite-dimensional and
finite-dimensional APIs are not aligned.  Although the user interacts with only
one of these APIs at any given time, an aligned API structure exposes the
opportunity for algorithms designed on function space, which tend to be more
robust algorithms, to be used in the finite-dimensional setting.  Moreover, an
aligned API eases the maintenance, documentation and testing burden.

Currently, there are only two (finite-dimensional) algorithms the user can use,
DRAM and Multi-level.  At present, there is no organized structure that
Markov chains (\texttt{MetropolisHastingsSG} objects) inherit from, meaning
that there is a significant hurdle involved in bootstrapping one's own MCMC
algorithm for the purposes of testing and research.  Just as above, a
consistent class hierarchy for MCMC algorithms would ease the burden for software
maintenance.

A rather cumbersome design choice made early on in the development of \Queso\
was the hot-swappability of vector and matrix implementations for all of
\Queso's classes.  The net result of this is that any \Queso\ class that
involves an operation with a vector or a matrix is templated around the type of
that vector or matrix.  This was done to ensure that optimsed code could be
generated that dealt with the specifics of each vector and matrix library.
Assuming that, in high-performance uncertainty quantification, likelihood
evaluations are the dominating cost of Markov-chain Monte Carlo sampling, one
need not encumber the \Queso API with such templates.  Furthermore, a
hierarchical class structure for vector and matrix types exists in \Queso\ and
therefore necessitates the run-time overhead of virtual table lookups.  Efforts
are currently ongoing to enrich the vector and matrix class hierarchy in
\Queso\ sufficiently such that the particulars of vector and matrix
implementations still remain opaque, but significantly shorten unnecessarily
long class names with a negligibly small impact on run-time performance.  This
enrichment would also allow \Queso\ to pick a high-tuned vector/matrix
implementation at configure-time for high-performance problems in exascale
compute environments.  For example, \Queso's build system could default to
using PETsc vectors optimized for multi-core architectures while the user need
not deal explicitly with \MPI\ calls.  All parallel logic would be handled under
the hood.  This offers a pleasing software experience while maintaining
performance.

Python has become a very popular environment to do post-processing and
visualization in multi-core HPC systems.  A desirable feature to have in
\Queso\ would be the automatic generation of python bindings.  This would offer
the possibility to do uncertainty quantification in statistical inverse
problems as a quick-turnaround experiment for cheap forward models in an
interpreted language environment. This implementation will likely
leverage the Simplified Wrapper and Interface Generator (SWIG) which is
not limited to Python, and can provide interfaces to many modern
programming languages, such as Perl, Python, Ruby, and Tcl.

%

\subsection{Exascale}

Uncertainty quantification has pushed the limits of current computational power
by requiring many evaluations of large-scale engineering systems described by
partial differential equations.  Utilizing more information about the system
can significantly increase the performance of MCMC algorithms
\cite{Girolami2011,Martin2012,Bui-thanh2012}.
In particular \Queso\ does not currently implement MCMC algorithms that use
gradient- or Hessian-based to inform proposal distributions.  However, the
design of the API for the pure virtual methods in \texttt{BaseScalarFunction}
allow this information to be passed to \Queso\ easily, in the form of a pointer
\texttt{V * gradVector}.  For more details on the parameters passed to the
\texttt{lnValue} function, the reader is directed to the \Queso\ documentation
which be found online here:  \url{http://libqueso.com}.


%

\bibliography{refs}
\bibliographystyle{plain}

\end{document}